%% file: draft.tex
\documentclass[lettersize,journal]{IEEEtran}
\usepackage{amsmath,amsfonts}
\usepackage{algorithmic}
\usepackage{algorithm}
\usepackage{amssymb}
\usepackage{array}
\usepackage[caption=false,font=normalsize,labelfont=sf,textfont=sf]{subfig}
\usepackage{textcomp}
\usepackage{stfloats}
\usepackage{url}
\usepackage{microtype}
\usepackage{verbatim}
\usepackage{graphicx}
\usepackage{cite}

\usepackage{amssymb}    
\usepackage[utf8]{inputenc} 

\usepackage{hyperref}   
\usepackage{xcolor}      
\usepackage{comment}    
\usepackage{soul}     
\usepackage{capt-of} 
\usepackage{etoolbox} 

\graphicspath{{figures/}} 

\newcommand{\ourname}[1][]{LCIP\ifx&#1&\else\textsubscript{#1}\fi}
\newcommand{\citep}[1]{\cite{#1}} 
\newcommand{\citet}[1]{\cite{#1}} 
\newcommand{\citeyearp}[1]{\cite{#1}} 

\begin{document}

\title{LCIP: Loss-Controlled Inverse Projection of High-Dimensional Image Data}
\author{Yu Wang, Frederik L. Dennig, Michael Behrisch, Alexandru Telea        
\thanks{Yu Wang, Michael Behrisch, and Alexandru Telea are with Utrecht University, The Netherlands. Email: yuwang@duck.com, m.behrisch@uu.nl, a.c.telea@uu.nl}
\thanks{Frederik L. Dennig is with University of Konstanz, Germany. Email: frederik.dennig@uni-konstanz.de}} 

\makeatletter
\apptocmd{\@maketitle}{\centering%
    \setcounter{figure}{0} 
    \includegraphics[width=0.8\linewidth]{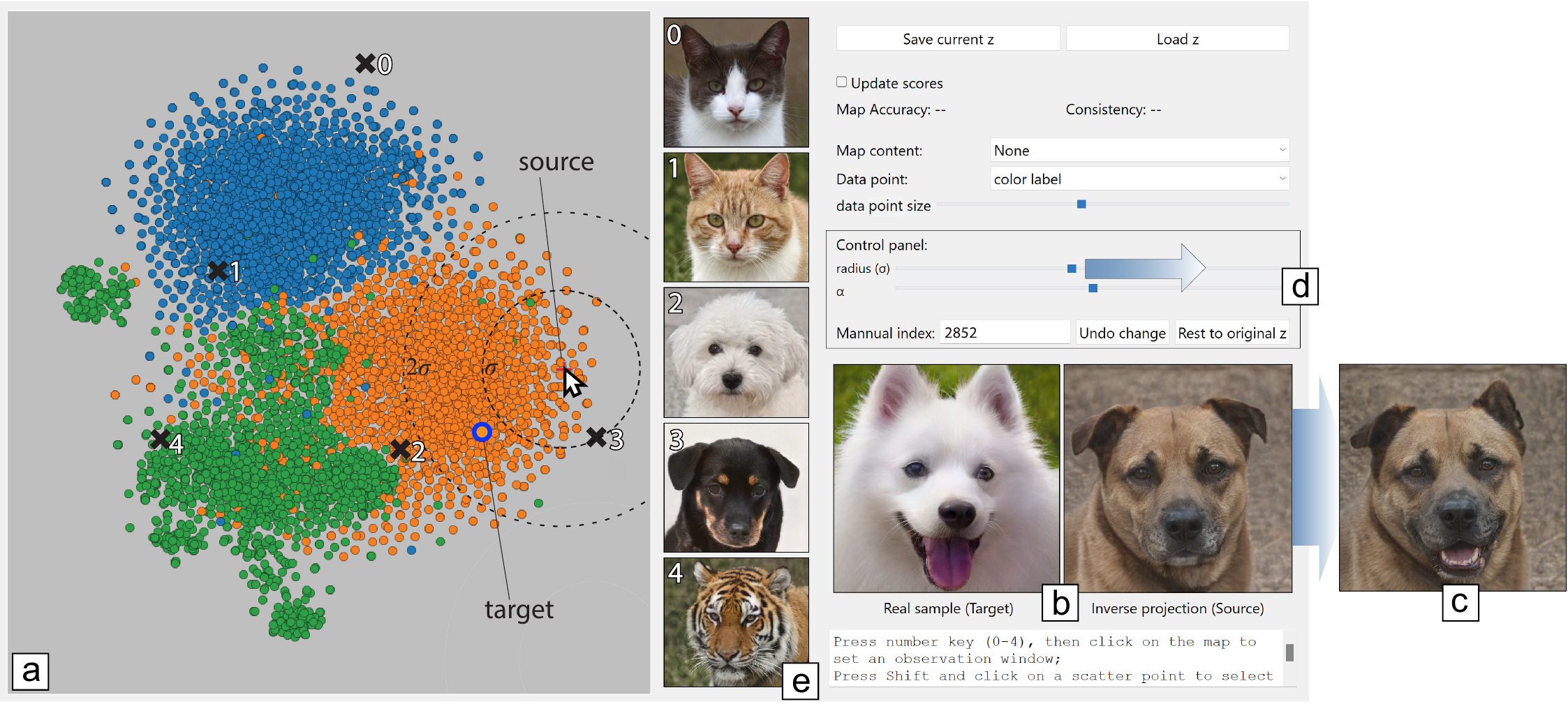}
    \captionof{figure}{\protect\input{sections/teaser_caption}}
    \label{fig:teaser}
    \vspace{-0.15cm}
}{}{}
\makeatother

\markboth{Journal of \LaTeX\ Class Files,~Vol.~14, No.~8, August~2021}%
{Shell \MakeLowercase{\textit{et al.}}: A Sample Article Using IEEEtran.cls for IEEE Journals}

\maketitle

\begin{abstract}
Projections (or dimensionality reduction) methods $P$ aim to map high-dimensional data to typically 2D scatterplots for visual exploration. Inverse projection methods $P^{-1}$ aim to map this 2D space to the data space to support tasks such as data augmentation, classifier analysis, and data imputation. Current $P^{-1}$ methods suffer from a fundamental limitation -- they can only generate a fixed surface-like structure in data space, which poorly covers the richness of this space. We address this by a new method that can `sweep' the data space under user control. Our method works generically for any $P$ technique and dataset, is controlled by two intuitive user-set parameters, and is simple to implement. We demonstrate it by an extensive application involving image manipulation for style transfer.
\end{abstract}


\input{main_content}



\bibliographystyle{IEEEtran}
\bibliography{refs} 





\end{document}

%% file: sections/teaser_caption.tex
LCIP illustrated by a style transfer application. 1) Brush the projection (a) to find a frowning dog which we want to make smile. The image of this dog is shown in (b, right). 2) Use the same brushing to find a target image with a smiling face (b, left). 3) Pulling a slider (d) changes the source towards the target – that is, makes the frowning dog (b) smile like the target, without changing its identity, see result in (c). Optionally, we can adjust the influence range $\sigma$ to affect how other images close to the source also get changed towards the target – see the circles centered at mouse location in (a). We can also select a few additional points in the projection – see locations marked 0-4 in (a) – and inspect how the user-controlled inverse projection behaves there. Point 3 is close to the source, so it will be strongly influenced by the source’s change; points 0, 1, and 4 are far away from the source, so they will not be affected.

%% file: main_content.tex
\section{Introduction}
\label{sec:introduction}

\input{sections/intro1.tex}

\input{sections/intro2.tex}

\input{sections/intro3.tex}

\section{Background and Related Work} 
\label{sec:related_work}

\input{sections/related1_definition.tex}

\input{sections/related2_pinv.tex}

\input{sections/related3_summary_pinv.tex}
\input{sections/related4_disentangle.tex}

\input{sections/method.tex}

\input{sections/evaluation.tex}

\input{sections/application.tex}

\input{sections/user-study}

\input{sections/disscussion.tex}

\input{sections/conclusion.tex}

%% file: sections/intro1.tex
\IEEEPARstart{M}{ultidimensional} projections, also called dimensionality reduction methods, are established techniques for the visual exploration of high-dimensional data\,\citep{lespinats11,sorzano2014surveydimensionality,nonato18}. Many DR methods have been proposed in the past decades, such as PCA, t-SNE, and UMAP, to mention just a few.

\emph{Inverse projections} aim to revert the mapping produced by a given projection. They aim to generate highly plausible, yet hypothetical, data items from any location in the embedding space created by projecting a given dataset. This enables applications such as data augmentation\,\citep{rodrigues2020VisualAnalytics,benato2024HumanloopUsing}, analyzing trained ML classification models by so-called decision maps\,\citep{DBM2019rodrigues,oliveiraSDBM2022,schulz2020DeepViewVisualizing}, pseudolabeling for creating training sets\,\citep{benato2024HumanloopUsing}, morphing and data imputation\,\citep{Amorim2012ilamp,espadoto2021unprojection}, and testing a classifier's brittleness \emph{vs} backdoor and data poisoning attacks\,\citep{schulz2020DeepViewVisualizing,differentiableDBM}. 

Both direct and inverse projections suffer from significant \emph{information loss} \citep{zheng2023AutoencodersIntrinsic}. Direct projections cannot, in general, map data whose intrinsic dimensionality largely exceeds that of the 2D projection space while preserving data structure, \emph{i.e.}, inter-point distances and neighborhoods\,\citep{nonato18,espadoto19}. Yet, users expect that inverse projections allow them to examine large parts of a data space starting from its 2D projection space. However, a recent study has shown that all current inverse projection techniques only generate fixed, surface-like, structures in the data space, so the \emph{coverage} of inverse projections is limited to a small part of the data\,\citep{wang2024FundamentalLimitations}.

%% file: sections/intro2.tex
In this work, we enable inverse projections to represent surface-like structures whose location in data space is not fixed but controllable by the user. Intuitively put, the user controls two parameters that allow the surface to move between a fixed position -- given by the direct projection -- and positions that get close to data samples deemed of interest by the user. As one changes the control parameters, the said surface literally `sweeps' the data space, thereby covering points which a fixed inverse projection, computed by existing techniques, could not reach. Specifically, for a sample $\mathbf{x}$ projecting to $\mathbf{y}=P(\mathbf{x})$ by a user-chosen technique $P$, we find the information $\mathbf{z}$ that is lost by $P$ and enable users to (interactively) control how to combine $\mathbf{y}$ (information preserved by $P$) and $\mathbf{z}$ (information lost by $P$) to compute our controllable inverse projection.

%% file: sections/intro3.tex
To realize this, we must answer three questions: (1) How to ensure $\mathbf{z}$ is \emph{independent} of $\mathbf{y}$ (as we want to let $P$ compute $\mathbf{y}$ and the user to control $\mathbf{z}$)? (2) How to compute $\mathbf{z}$ for locations in the 2D space where \emph{no ground-truth} sample $\mathbf{x}$ projects? (3) How to \emph{control} the structure created by the inverse projection, to enable users to explore large parts of the data space?

Our Loss-Controlled Inverse Projection (LCIP) answers these questions as follows: (1) We minimize mutual information to disentangle the $\mathbf{y}$ and $\mathbf{z}$ spaces. (2) We use interpolation between $\mathbf{z}$ values learned for known data samples to fill in the `empty' areas in a projection. (3) We allow users to interactively adapt $\mathbf{z}$ to control the shape of the inverse projection. 

Summarizing our contribution, we address the fundamental limitations of inverse projection techniques -- limited and fixed coverage.  To show our method's abilities, we first study its disentanglement effect (Sec.~\ref{sec:disentanglement}) and next compare LCIP both visually and via quality metrics with 3 state-of-the-art inverse projections for 4 datasets and 2 direct projections (Sec.~\ref{sec:compare_previous}). Next, we show how LCIP's interactive control creates higher-dimensional structures than surfaces, something existing methods cannot achieve (Sec.~\ref{sec:control_and_ID}). Finally, we show how our method can be used to support a style transfer application (Sec.~\ref{sec:application_afhq}). While our method can technically handle arbitrary high-dimensional data, we focus on image data which is easily interpretable -- an important requirement for our technique where users aim to interactively control the structure created by the inverse projection.

%% file: sections/related1_definition.tex
\subsection{Projections and inverse projections}
\label{sec:basics}
\noindent\textbf{Projection:} Let $X=\{\mathbf{x}_i\}$, $\mathbf{x}_i \in \mathbb{R}^n$, $1 \leq i \leq N$, be an $n$-dimensional dataset. A projection 
\begin{equation}
  P (X) = \{ \mathbf{y}_i \} \subset \mathbb{R}^q
\end{equation}
maps each $\mathbf{x}_i \in X$ to a $\mathbf{y}_i \in \mathbb{R}^q$, where $q \ll n$. For visualization goals as in this work, we typically set $q=2$. Projection techniques, such as PCA, t-SNE\,\citep{tSNEMaaten2008} and UMAP\,\citep{UMAP2018}, are covered by extensive surveys\,\citeyearp{espadoto19,nonato18}. We further denote any point in the projection space $\mathbb{R}^2$ by $\mathbf{p}$. 

\smallskip
\noindent \textbf{Inverse projection:} An inverse projection, also called backprojection or unprojection, is a function
\begin{equation}
  P^{-1}: \mathbb{R}^q \rightarrow \mathbb{R}^n
\end{equation}
that aims to reverse the effect of a given $P$. We further denote by $\mathbf{q}$ points in the data space obtained by applying the inverse projection to some 2D point $\mathbf{p}$, \emph{i.e.}, $\mathbf{q} = P^{-1}(\mathbf{p}).$ For a dataset $X$ and its projection $Y=P(X)$,  $P^{-1}$ is typically computed by minimizing errors of the form $\| P^{-1}(P(\mathbf{x}_i)) - \mathbf{x}_i\|$ over $X$. Crucially for applications, once $P^{-1}$ is defined based on $X$ and $Y$, one can use it to infer data values for \emph{any} $\mathbf{p} \in \mathbb{R}^2$, including so-called `gap areas' $\mathbb{R}^2 \setminus Y$ between points in $Y$.

Unlike direct projections, only a few inverse projection methods exist. iLAMP\,\citep{Amorim2012ilamp} builds local affine mappings that revert the LAMP direct projection\,\cite{lamp}. To address the lack of continuity and global mapping in iLAMP, a method using Radial Basis Functions (RBFs) was next proposed,\,\citep{amorim2015Facinghighdimensions}. UMAP\,\citep{UMAP2018} constructs both a direct and inverse projection based on preserving the data structure's topology from $\mathbb{R}^n$ to $\mathbb{R}^q$. DeepView\,\citep{schulz2020DeepViewVisualizing} customized UMAP with a classifier to produce a discriminative projection; a modified UMAP was used for inverse projection. Self-Supervised Network Projection (SSNP)\,\citep{espadoto2021SSNP} enhanced standard autoencoders\,\cite{hinton2006reducing} with self-supervision using pseudolabels to produce both direct and inverse projections. Blumberg \emph{et al.}\,\citep{invertingMDS2024EuroVA} proposed a method to invert multidimensional scaling (MDS) that works well for $q < 10$\,\citep{torgerson52,kruskal1964multidimensional}. Finally, NNinv\,\citep{espadoto2019NNinv,espadoto2021unprojection} used supervised deep learning to invert any projection $P(X)$ in a generic way, a method that was recently extended to autoencoders\,\cite{dennig2025Evaluating}.

%% file: sections/related2_pinv.tex
\vspace{0.15cm}
\noindent \textbf{Inverse projection applications:} Inverse projections allow performing dynamic imputation to explore a high-dimensional data space using its projection\,\citep{espadoto2021unprojection}; Amorim \emph{et al.} morphed between facial expressions using RBF inverse projections\,\citep{amorim2015Facinghighdimensions}. Inverse projections enable the creation of so-called \emph{decision maps}, \emph{i.e.}, 2D images that capture the behavior of a trained ML classification model. Such maps provide insights into a model's behavior\,\citep{oliveiraSDBM2022,schulz2020DeepViewVisualizing}; allow pseudolabeling samples to create rich training sets\,\citep{benato2024HumanloopUsing}; and support explorations such as studying a model's brittleness against data attacks\,\citep{differentiableDBM}. 
In all such applications, inverse projections are a key element of \emph{interactive visualization} applications: In \cite{amorim2015Facinghighdimensions}, users interactively sweep the projection space to create 3D models that morph between those present in a given set. In decision map applications, users interactively brush the decision map (created using inverse projections) to find which types of samples project \emph{e.g.} in areas close to decision boundaries or in other zones where misclassifications can occur. 
An extension of this scenario allows users to locate wrongly-classified zones in decision maps and interactively select samples falling in such areas to improve classification performance\,\cite{benato2024HumanloopUsing}. In our work, we extend this interactivity by allowing users to actually determine the shape of the inverse projection.

\vspace{0.15cm}
\noindent \textbf{Quality of inverse projections:} Quality of inverse projections can be measured by their MSE $\| P^{-1}(P(\mathbf{x})) - \mathbf{x}\|$. However, this can only be done for points in $X$; for the so-called gap points, where inverse projections are actually useful, we do not have ground truth to test against. Previous work\,\citep{amorim2015Facinghighdimensions,espadoto2021unprojection,wang2023DMcompare,differentiableDBM} shows that \emph{smooth} inverse projections are overall preferred as they ensure that small changes in 2D space (\emph{e.g.}, when the user drags a 2D point to unproject) yield small changes in data space. Gradient maps\,\citep{espadoto2021unprojection} gauge this smoothness by computing the norm of the gradient of $P^{-1}$. Additional metrics capture the quality of decision maps\,\citep{wang2023DMcompare,differentiableDBM} which rely on -- but are not the same as -- inverse projections. We will use both MSE and gradient maps to evaluate our proposal.

\vspace{0.15cm}
\noindent \textbf{Surface limitation of inverse projections:} Recent work\,\citep{wang2024FundamentalLimitations} showed that all existing inverse projections map the 2D space to a \emph{fixed}, \emph{surface-like}, structure embedded in data space -- that is, a structure with intrinsic dimensionality close to two\,\citep{ansuini2019Intrinsicdimension}. While not unexpected, given that an inverse projection aims to smoothly map $\mathbb{R}^2$ to $\mathbb{R}^n$, this means that inverse projections can only cover a very \emph{limited subspace} of the data space $\mathbb{R}^n$. Yet, this subspace is completely non-transparent to, and not controllable by, the users. 
The above imply serious limitations for inverse projection applications. Using inverse projections for data augmentation will lack diversity. Separately, how can one be sure that a decision map truly depicts a classifier's behavior when it visualizes only a small part of the data space -- more specifically, the intersections of the true decision boundaries with the surface-like structure created by $P^{-1}$\,\citep{wang2024FundamentalLimitations}? How to explore the data space \emph{outside} this surface-like structure, and how to \emph{control} this exploration, are two challenges not answered by existing methods. In our work, we enable users to interactively control how such surface-like structures sweep the data space by using simple visual controls.

%% file: sections/related3_summary_pinv.tex
\vspace{0.15cm}
\noindent\textbf{Information loss:} Direct and inverse projections share limitations \emph{vs} information loss. For direct projections, if the data $X$ do not land on or close to a manifold, it is hard to construct a mapping $P(X)$ that fully preserves the data structure\,\citep{zheng2023AutoencodersIntrinsic}. For inverse projections, the limitations are stronger -- these always create a surface-like structure when mapping the 2D space to data space\,\cite{wang2024FundamentalLimitations}. Given the above, if we consider the project-unproject cycle, \emph{information loss} always occurs.

Inverse projections are structurally similar to data reconstruction or data generation tasks which aim to output high-dimensional data from low-dimensional representations. From this perspective, the ($P$, $P^{-1}$) cycle can be seen as an encoder-decoder structure, where the bottleneck is the 2D latent space. The dimensionality of this space is a critical factor for the quality of reconstruction and generation\,\citep{wang2016Autoencoderbased,marin2021effectlatent,padala2021EffectInput}. Inspired by these observations, we aim to break the limitations imposed by our bottleneck -- the visual space -- by retrieving information lost during $P$ and using it to drive the construction of $P^{-1}$ under user control. This will enable our $P^{-1}$ to span structures with higher intrinsic dimensionality than two, and also control where these are placed in data space.

%% file: sections/related4_disentangle.tex
\subsection{Disentangled representations and adversarial training}
\label{sec:gan}
As Sec.~\ref{sec:introduction} stated, our first goal is to find the information $\mathbf{z}$ lost during projection independent of the projection $\mathbf{y}$.
Independence is often relaxed to minimizing mutual information or separating complementary factors\,\citep{moyer2018InvariantRepresentations}. When achieved, this improves interpretability, reduces potential bias, and enhances generalization. For example, in handwriting recognition, separating text content $\mathbf{y}$ (what an actual letter \emph{is}) from its style $\mathbf{z}$ (how the letter is \emph{written}) helps model generalization. In speech processing, one aims to separate the speech content $\mathbf{y}$ from the speaker's identity $\mathbf{z}$\,\citep{hadad2018twostepdisentanglement}.

Minimizing mutual information between two latent representations, also called learning disentangled representations, can be done via adversarial training\,\citep{mathieu2016Disentanglingfactors,hadad2018twostepdisentanglement,xie2017ControllableInvariance,jaiswal2019UnifiedAdversarial,zheng2019DisentanglingLatent}.
Adversarial training, first used by generative adversarial networks (GANs) for image generation\,\citep{goodfellow2014GenerativeAdversarial}, can generate high-dimensional realistic samples from low-dimensional latent codes. GANs jointly train a generator $G$ and a discriminator $Dis$; $G$ aims to create samples that are indistinguishable from real samples; $Dis$ tries to distinguish real from generated samples. Adversarial training has been extended to other areas like robust machine learning, domain adaptation, and disentangled representation learning. While diffusion models are now more popular for image generation, GANs are significantly more efficient\,\citep{pan2023DragYour}. For example, DragGAN\,\citep{pan2023DragYour}, an interactive image manipulation method, uses the StyleGAN2 architecture\,\citep{karras2020AnalyzingImproving}.

Directly projecting high-dimensional datasets such as high-resolution images is challenging. In such cases, one typically describes the data using the latent space of a pre-trained classifier such as InceptionV3 or VGG16\,\citep{espadoto2020Deeplearning,benato2024HumanloopUsing}. Since we focus on inverse projection, we prefer to retrieve images from the latent space of a generative model designed for this purpose such as StyleGAN2. In more detail: Let $\mathbf{w} \in \mathcal{W}$ be the latent code that StyleGAN2 uses to generate images. Codes $\mathbf{w}$ can be obtained by inverting StyleGAN2 pre-trained on the same dataset\,\citep{karras2020AnalyzingImproving}. For a given $\mathbf{w}$, the corresponding image is then given by $G(\mathbf{w})$.  Adversarial training connects to our work in two ways: (1) We use it to enforce disentanglement between the information $\mathbf{z}$ and the projection $\mathbf{y}$ during the training of our inverse projection. (2) We use the $\mathcal{W}$ space of an image dataset to ease the projection process.

%% file: sections/method.tex
\section{Design of Loss-Controlled Inverse Projection}
\label{sec:method}
%

\subsection{Inverse Projection Deep Learning Network Architecture}
\label{sec:design}
\begin{figure}[!htbp]
  \centering
  \includegraphics[width=0.8\linewidth]{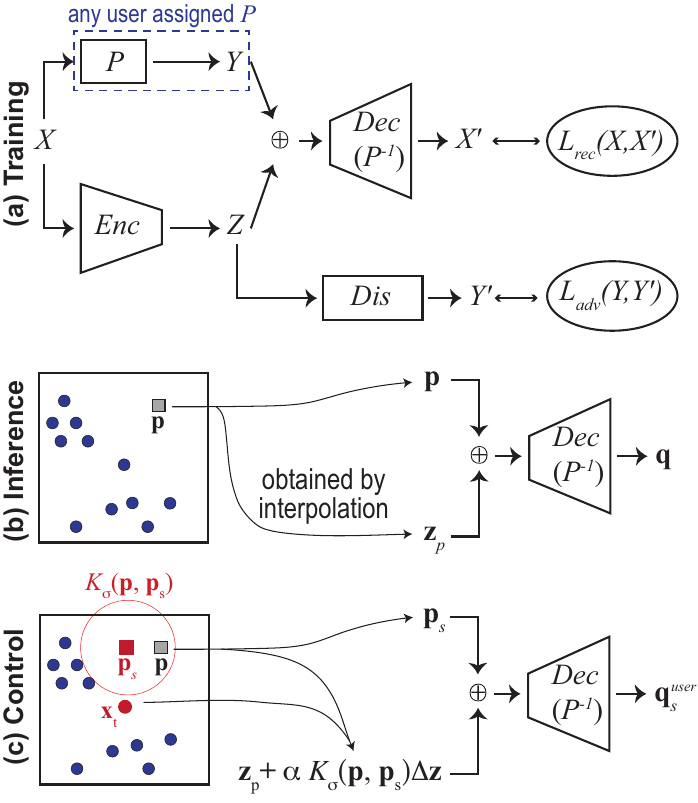}
  \caption{LCIP workflow. $\oplus$ denotes concatenation. (a) Training $Dis$ and $Dec$. $P$ is a user-selected DR method that projects $X$ to $Y$. $Enc$ encodes $X$ into $Z$. $Dis$ uses $Z$ to predict $Y$. $Dec$ ($P^{-1}$) uses $Y$ and $Z$ to reconstruct $X$. The adversarial network $Dis$ enforces disentanglement (Sec.~\ref{sec:design}). (b) Inversely projecting a 2D point $\mathbf{p}$ to data sample $\mathbf{q}$ (Sec.~\ref{sec:initialize_z}). (c) Users can  refine the inverse projection by maneuvering the controls marked in red: a source point $\mathbf{p}_s$, a target point $\mathbf{x}_t$, a pull factor $\alpha$, and a kernel radius $\sigma$. (Sec.~\ref{sec:control_mechanism}).}
  \vspace{-0.15cm}
  \label{fig:framework}
\end{figure}

We implement our inverse projection using neural networks. Consider a dataset $X$ and its projection $Y = P(X)$ computed by any user-chosen projection technique $P$.
Our network has two key parts: an encoder $Enc$ that computes the information in $X$ lost by $P$; and a decoder $Dec$ which is also our inverse projection $P^{-1}$. $Enc$ reads the data $X$ and outputs a latent code $Z = Enc(X) = \{\mathbf{z}_i\}$. $Dec$ reads the concatenation $Y \oplus Z$ and outputs the inversely projected data $X'  = Dec(Y, Z)$ (Fig.~\ref{fig:framework}a). We also need to ensure that $Z$ is not related to $Y$ (see Sec.~\ref{sec:gan}). We do this by a third network $Dis$ which reads $Z$ and outputs $Y'=Dis(Z)$ and is adversarially trained to minimize the error between $Y$ and $Y'$. $Enc$ and $Dec$ are jointly trained to (i) minimize the reconstruction error between $X$ and $X'$, and (ii) maximize the difference between $Y$ and $Y'$. 

Let $\theta_{Enc}$, $\theta_{Dec}$, $\theta_{Dis}$ be the weight and bias parameters of $Enc$, $Dec$, and $Dis$, respectively. 
Let $L_{adv}(Y, Y')$ be the reconstruction loss between $Y$ and $Y'$, and $L_{rec}(X, X')$ the reconstruction loss between $X$ and $X'$. 
When optimizing $\theta_{Dis}$, $L_{adv}$ is minimized, so \(Dis\) learns to predict \(Y\) from \(Z\).
The cost function $J$ for optimizing $Enc$ and $Dec$ is defined as 
\begin{equation}
  J = L_{rec}(X, X') - \lambda   L_{adv}(Y, Y'),
  \label{eqn:j}
\end{equation}
where $\lambda > 0$ is a hyperparameter that balances reconstruction \emph{vs} adversarial loss, with the target of the optimization being
\begin{equation}
  \min_{\theta_{Enc}, \theta_{Dec}} J,
\end{equation}
that is, we minimize $L_{rec}$  and maximize $L_{adv}$ while keeping $\theta_{Dis}$ fixed. 
Once trained, $Enc$ infers $Z$ from $X$ and $Dec$ next inversely projects $Y \oplus Z$ to $X'$.

\smallskip
\noindent\textbf{Implementation:} We use fully-connected networks for $Enc$, $Dec$, and $Dis$. $Enc$ has 3 hidden layers (sizes 512, 256, and 128). $Dec$ has 4 hidden layers (sizes 128, 256, 512, and 1024). Each hidden layer of $Enc$ and $Dec$ is followed by a ReLU activation function. The final layer of $Dec$ is followed by a sigmoid activation function. $Dis$ has 2 hidden layers, each with a size of 128. Each hidden layer is followed by batch normalization and a ReLU activation function. The dimension of $Z$ is set to 16 -- a value we empirically found to be sufficient to capture the information loss of all studied $P$ techniques. We use these settings consistently for all tested datasets.

We use mean squared error (MSE) for both $L_{rec}$ and $L_{adv}$ (Eqn.~\ref{eqn:j}). For $\lambda$ (Eqn.~\ref{eqn:j}), we ran a grid search over the range $[0.005, 4]$, and found that $\lambda \in [0.01, 0.1]$ gives good results for all tested datasets. We train all networks using the Adam optimizer with a learning rate of $0.001$. While training $Dis$, we have noticed that the adversarial training requires more steps to stabilize, since it learns from a changing input. Hence, at each iteration, we update $Dis$ 5 times, then update $Enc$ and $Dec$ once. Our work is implemented using PyTorch\,\citep{paszke2019PyTorch} with PySide6 (Qt) for the GUI, and is publicly available\,\citep{softwareLCIP}.

\subsection{Computing $\mathbf{z}$ for the entire projection space}
\label{sec:initialize_z}
To apply our inverse projection (Sec.~\ref{sec:design}) at a 2D projection point $\mathbf{y}_i$, we need the latent code $\mathbf{z}_i = Enc(\mathbf{x}_i)$ computed from the data sample $\mathbf{x}_i \in X$ that projects to $\mathbf{y}_i$. To apply our method to \emph{any} 2D point $\mathbf{p}$, we need to estimate $\mathbf{z}_\mathbf{p}$ at that location. We do this by interpolating the $\mathbf{z}_i$ values of the samples $X$ (see Fig.~\ref{fig:framework}b). We tested two methods: weighted k-NN ($k=10$ neighbors) and smoothed RBF with a parameter-free thin plate spline kernel. RBF gives a smooth surface, while weighted k-NN is slightly faster. We discuss the results of both interpolation methods in Sec.~\ref{sec:compare_previous}. 

Having now the latent code $\mathbf{z}_\mathbf{p}$ for any $\mathbf{p} \in \mathbb{R}^2$, we can inversely project $\mathbf{p}$ to the data space (Fig.~\ref{fig:framework}b) as 
\begin{equation}
\mathbf{q} = P^{-1}(\mathbf{p}) = Dec(\mathbf{p}, \mathbf{z}_\mathbf{p}).
\label{eq:invproj}
\end{equation}

\subsection{Controlling the inverse projection}
\label{sec:control_mechanism}
To allow users to effectively control the shape of the inverse projection in data space, two questions arise: (1) How to do this \emph{easily}, \emph{i.e.}, by changing a small number of intuitive parameters 
in a direct, visual way; and (2) How to make our $P^{-1}$ cover zones in data space where \emph{plausible} samples exist, so that $P^{-1}$ is useful for real-world applications.

\begin{figure}[!htbp]
  \centering
  \includegraphics[width=1.0\linewidth]{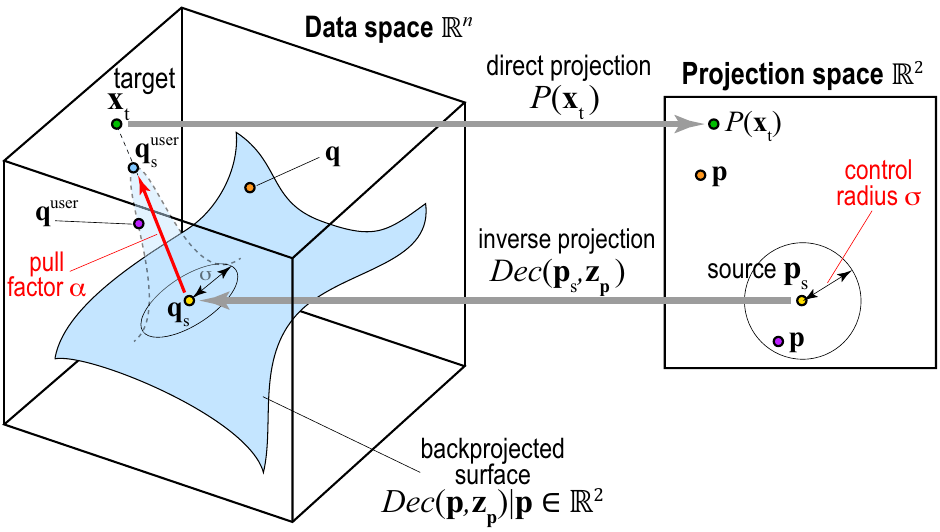}
  \caption{Controlling the inverse projection. User parameters are marked in red.}
  \label{fig:control}
  \vspace{-0.15cm}
\end{figure}

We solve both problems by the pipeline shown in Fig.~\ref{fig:framework}c, which we explain next (see also Fig.~\ref{fig:control}). Let $\mathbf{x}_t \in X$, called a \emph{target} sample, be a data point that we want to make our inverse projection go close to. Users can discover such points by brushing the projection with a tooltip to see the values $\mathbf{x}_t$ (Fig.~\ref{fig:control}, green point). The user next selects a 2D \emph{source} point $\mathbf{p}_s$ and manipulates it to control $P^{-1}$ (Fig.~\ref{fig:control}, yellow point). We then `pull' the inverse projection $P^{-1}(\mathbf{p}_s)$ of $\mathbf{p}_s$ towards $\mathbf{x}_t$  by adjusting $\mathbf{z}_{\mathbf{p}_s}$ (initially computed by interpolation as described in Sec.~\ref{sec:initialize_z}) -- see the red arrow in Fig.~\ref{fig:control}. To achieve this pull, we first compute the difference 
\begin{equation}
  \label{eq:delta_z}
  \begin{split}
    \Delta \mathbf{z} &= \mathbf{z}_t - \mathbf{z}_{\mathbf{p}_s} = Enc(\mathbf{x}_t) - \mathbf{z}_{\mathbf{p}_s}.
  \end{split}
\end{equation}
Intuitively put, $\Delta z$ tells how much the latent codes of the source and target differ -- that is, what we need to change in the source's inverse projection  to make it become the target. With $\Delta \mathbf{z}$, we compute the user-controlled inverse projection of $\mathbf{p}_s$ as 
\begin{equation}
  \label{eq:control_single}
  \mathbf{q}_s^{user} = Dec(\mathbf{p}_s, \mathbf{z}_{\mathbf{p}_s} + \alpha  \Delta \mathbf{z}),
\end{equation}
where $\alpha \in \mathbb{R}$ is a factor giving the pull magnitude (Fig.~\ref{fig:control}, light blue point).
Yet, this adjustment only changes $P^{-1}$ at the \emph{single} location $\mathbf{q}_s^{user}$. 2D points close to $\mathbf{p}_s$ will not be affected by this pull, as their inverse projections still follow Eqn.~\ref{eq:invproj}. The inverse projection will exhibit a discontinuity or lack of smoothness around $\mathbf{q}_s^{user}$, which is undesired (Sec.~\ref{sec:basics}). We could get smoothness by applying $\Delta \mathbf{z}$ to all such 2D points. However, this changes the inverse projection \emph{globally} -- the source $\mathbf{p}_s$ will equally influence all inversely-projected points, no matter how far these are from $\mathbf{p}_s$. We jointly achieve smoothness and local control by weighing the adjustment based on distances in the projection space to the source $\mathbf{p}_s$. That is, after adjusting the inverse projection at $\mathbf{p}_s$ (Eqn.~\ref{eq:control_single}), we replace Eqn.~\ref{eq:invproj} by
\begin{equation}
  \label{eq:control_kernel}
  \mathbf{q}^{user} = Dec(\mathbf{p}, \mathbf{z}_p + \alpha  K_{\sigma}(\mathbf{p}, \mathbf{p}_s)  \Delta \mathbf{z}),
\end{equation}
where $K_{\sigma}(\mathbf{p}, \mathbf{p}_s) = e^{-\frac{\|\mathbf{p} - \mathbf{p}_s\|^2}{2\sigma^2}}$ is a Gaussian centered at $\mathbf{p}_s$ and $\sigma$ controls the source's influence. Larger $\sigma$ values make the control more global and yield smoother inverse projections; smaller values have the opposite effect. When $\mathbf{p}$ is close to the source $\mathbf{p}_s$ (Fig.~\ref{fig:control} purple point), its inverse projection gets pulled towards the target $\mathbf{x}_t$ -- see light-blue bump on the surface in Fig.~\ref{fig:control}. When $\mathbf{p}$ is far from $\mathbf{p}_s$ (Fig.~\ref{fig:control} orange point), its inverse projection stays on the surface given by Eqn.~\ref{eq:invproj}.

Figure~\ref{fig:teaser} shows this control mechanism in action for a simple style transfer application. Here, the dataset $X$ contains images of various animal faces. The user selects the source $\mathbf{p}$ by picking a location in the projection -- not necessarily an actual projected sample. The image $\mathbf{q}$ for this point is computed by the inverse projection -- see the sad-looking dog in Fig.~\ref{fig:teaser}b, right. Next, the user selects a target sample $\mathbf{x}_t$ -- see the happy-looking dog in Fig.~\ref{fig:teaser}b, left. Pulling a slider changes $\alpha$ and morphs the source image towards the target. The user can see how far/strong the effect of changing the source propagates over the projection by selecting other images (Fig.~\ref{fig:teaser}e) and assessing their changes during source manipulation.

%% file: sections/evaluation.tex
\section{Evaluation}
\label{sec:experiments}
We evaluate our proposed inverse projection method on several datasets and projection techniques. We first study the effect of disentanglement both qualitatively and quantitatively and show that our latent codes $\mathbf{z}$ are indeed independent of the projected information $\mathbf{y}$ (Sec.~\ref{sec:disentanglement}). Then, we compare our inverse projection (without interactive control) to existing inverse projection methods and show we reach similar quality (Sec.~\ref{sec:compare_previous}). Finally, we show that our interactive control can break the surface-like limitation discussed in Sec.~\ref{sec:control_and_ID}.

For $P$ we use t-SNE and UMAP, two of the highest-quality techniques in DR landscape\,\citep{espadoto19}, also used in other inverse-projection studies\,\citep{espadoto2019NNinv,espadoto2021unprojection,schulz2020DeepViewVisualizing,wang2023DMcompare}. We use the following datasets:

\smallskip
\noindent\textbf{MNIST:} 70K samples of handwritten digits (0-9), each a $28^2$ grayscale image flattened to a 784-size vector\,\citep{lecun2010MNISThandwritten}.

\smallskip
\noindent\textbf{Fashion-MNIST:} 70K samples of 10 fashion categories (e.g., T-shirts, trousers, dresses), each a $28^2$ grayscale image flattened to a 784-size vector\,\citep{xiao2017FashionMNISTnovel}.

\smallskip
\noindent\textbf{HAR:} 10K samples of smartphone accelerometer and gyroscope data capturing six human activities (walking, walking upstairs, walking downstairs, sitting, standing, laying)\,\citep{anguita2012human}.

\smallskip
\noindent\textbf{$\mathcal{W}$ of AFHQv2:} AFHQv2 has 15K color images of animal faces in 3 classes: dogs, cats, wild animals\,\citep{choi2020StarGANv2}. Its $\mathcal{W}$ is a $\mathbb{R}^{512}$ latent space used by StyleGAN2 models\,\citep{karras2020AnalyzingImproving} to generate images (Sec.~\ref{sec:gan}). For ease of exposition, we next show the generated images $G(\mathbf{w})$ instead of the raw codes $\mathbf{w}$. 

For a dataset $X = \{\mathbf{x}_i\}$ and its projection $Y = \{\mathbf{y}_i\}$, $\mathbf{y}_i = P(\mathbf{x}_i)$, we form the paired set $D=\{(X, Y)\}$, and split it into training $D_T=\{(X_T, Y_T)\}$ and test $D_v=\{(X_v, Y_v)\}$ subsets.
Table~\ref{tab:datasets} summarizes the above, including the dataset-specific choice of the hyperparameter $\lambda$ (Eqn.~\ref{eqn:j}).

\begin{table}[h!]
  \centering
  \scriptsize
  \caption{Datasets used in our evaluation with dimensionality $n$, sample count $|X|$, training and testing set sizes $|D_T|$ and $|D_v|$, and values of hyperparameter $\lambda$ (Eqn.~\ref{eqn:j}).}
  \label{tab:datasets}
  \begin{tabular}{l|c|c|c|c|c}
  \hline
  \textbf{Dataset} & $n$ & $|D_T|$ & $|D_v|$ & $|X|$ & $\lambda$\\
  \hline
  MNIST & 784 & 5000 & 5000 & 70000 & $0.1$\\
  Fashion-MNIST & 784 & 5000 & 5000 & 70000 & $0.1$\\
  HAR & 561 & 5000 & 5000 & 10299 & $0.05$\\
  $\mathcal{W}$ of AFHQv2 & 512 & 5000 & 5000 & 14336 & $0.01$\\
  \hline
  \end{tabular}
\end{table}

\subsection{Added value of disentanglement}
\label{sec:disentanglement}
To show the added value of the disentanglement, we compare our results using the loss in Eqn.~\ref{eqn:j} (called next \emph{WithDis}) with the same network trained without $L_{adv}$ (called next \emph{NoDis}).

\smallskip
\noindent\textbf{Quantitative evaluation:} We measure disentanglement by how well $\mathbf{z}$ can predict $\mathbf{y}$. The \emph{worse} $\mathbf{z}$ can predict $\mathbf{y}$, the better the disentanglement, \emph{i.e.}, $\mathbf{z}$ and $\mathbf{y}$ are more independent\,\citep{jaiswal2018Unsupervisedadversarial,jaiswal2019UnifiedAdversarial,moyer2018InvariantRepresentations}.
We measure this by training a \emph{post-hoc} regression model to predict $Y_T$ from $Z_T=Enc(X_T)$. This model is a neural network with one hidden layer (100 units) followed by ReLU activation, trained for 200 epochs using the Adam optimizer. We measure $R^2$ and $MSE$ on a hold-out test set. We expect that 
\emph{WithDis} should yield low $R^2$ and high $MSE$ -- that is, $\mathbf{z}$ and $\mathbf{y}$ are independent and/or different. Conversely, we expect that \emph{NoDis} should yield high $R^2$ and low $MSE$ -- that is, $\mathbf{z}$ and $\mathbf{y}$ are correlated and/or similar. Table~\ref{tab:predict_y_from_z} shows that $MSE$ and $R^2$ for \emph{WithDis} and \emph{NoDis} indeed match the expectations, thus confirming our claimed disentanglement.

\begin{table}[!htb]  
  \centering
  \scriptsize
  \caption{Measuring disentanglement: How well can $\mathbf{z}$ predict $\mathbf{y}$?}
  \label{tab:predict_y_from_z}
  \begin{tabular}{ll|rr|rr}
  \hline
  \textbf{Dataset} & $P$    & \multicolumn{2}{c|}{\textbf{\emph{WithDis}}} & \multicolumn{2}{c}{\textbf{\emph{NoDis}}} \\
                     &               & $MSE$           & \textbf{$R^2$}          & $MSE$           & \textbf{$R^2$}           \\ \hline
    MNIST	& UMAP	& 17.1	& 0.032	& 1.7	& 0.900 \\
    MNIST	& t-SNE	& 1497.5	& 0.011	& 161.9	& 0.892 \\
    Fashion-MNIST	& UMAP	& 19.9	& 0.201	& 0.7	& 0.968 \\
    Fashion-MNIST	& t-SNE	& 1284.4	& 0.096	& 79.4	& 0.944 \\
    HAR	& UMAP	& 48.8	& 0.098	& 6.9	& 0.862 \\
    HAR	& t-SNE	& 1337.8	& 0.170	& 152.7	& 0.892 \\
    $\mathcal{W}$ of AFHQv2	& UMAP	& 5.8	& 0.014	& 3.6	& 0.402 \\
    $\mathcal{W}$ of AFHQv2	& t-SNE	& 664.8	& 0.024	& 336.2	& 0.500 \\
  \hline
  \end{tabular}
\end{table}

\begin{figure}[!htbp]
  \centering
  \includegraphics[width=1.0\linewidth]{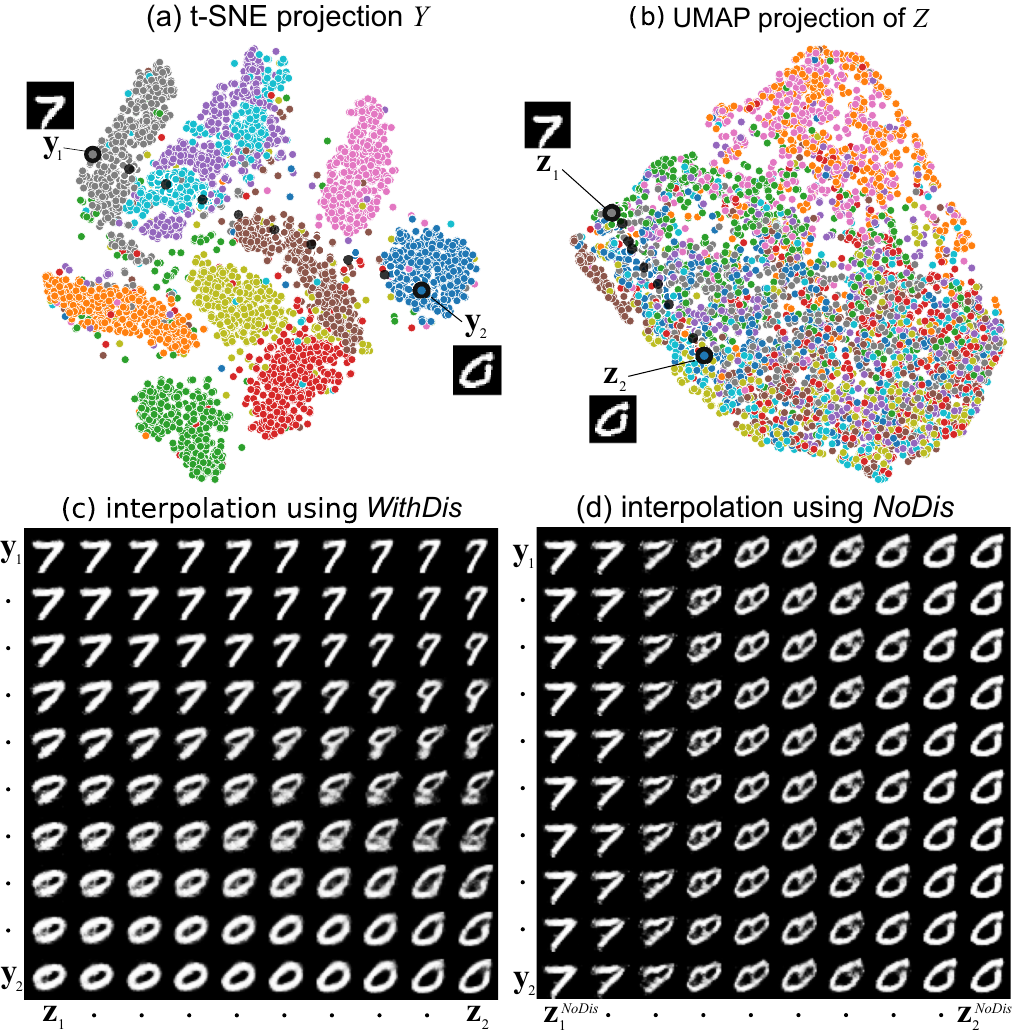}
  \caption{Showing disentanglement on the MNIST dataset. (a) 2D t-SNE projection $Y$. (b) UMAP projection of $Z$ using $WithDis$. (c) Inverse projections of the linear interpolation between two data points in the $Z$ and $Y$ spaces, using $WithDis$ (c) and $NoDis$ (d).}
  \label{fig:show_func_dis}
\end{figure}

\smallskip
\noindent\textbf{Qualitative evaluation:} 
We visually evaluate disentanglement by selecting two samples $\mathbf{x}_1$ and $\mathbf{x}_2$ of MNIST dataset and their t-SNE projections $\mathbf{y}_1$ and $\mathbf{y}_2$. These are two images of digits 0 and 7 -- see Fig.~\ref{fig:show_func_dis}a. Let $\mathbf{z}_1 = Enc(\mathbf{x}_1)$ and $\mathbf{z}_2 = Enc(\mathbf{x}_2)$ be the codes of these two images. These are shown in a UMAP projection of the $\mathbf{z}$ values for the dataset in Fig.~\ref{fig:show_func_dis}b.
We linearly interpolate between $\mathbf{y}_1$ and $\mathbf{y}_2$, and $\mathbf{z}_1$ and $\mathbf{z}_2$, respectively, with 10 steps. 
This yields $10^2$ interpolated values $(\mathbf{y}, \mathbf{z})$ which we  inversely project to the data space using both \emph{WithDis} (Fig.~\ref{fig:show_func_dis}c) and \emph{NoDis} (Fig.~\ref{fig:show_func_dis}d). We see from the t-SNE projection that the label information is well-preserved in the projection space (Fig.~\ref{fig:show_func_dis}a) -- that is, digits of the same \emph{type}, \emph{e.g.}, zeroes or sevens, are well grouped.
In contrast, the UMAP projection of $\mathbf{z}$ strongly mixes labels (Fig.~\ref{fig:show_func_dis}b). This is desired, as $\mathbf{z}$ should capture a digit's writing style and not its class. We next see that, as we change $\mathbf{z}$ using \emph{WithDis}, the digit's \emph{style}  changes -- see columns in Fig.~\ref{fig:show_func_dis}c; while, as we change $\mathbf{y}$, the digit \emph{itself} changes -- see rows in Fig.~\ref{fig:show_func_dis}c. Hence, \emph{WithDis} disentangles $\mathbf{y}$ and $\mathbf{z}$ well. In contrast, for \emph{NoDis}, the inverse projection changes only when $\mathbf{z}$ changes -- see columns in Fig.~\ref{fig:show_func_dis}d; when $\mathbf{y}$ changes, the digit stays the same -- see rows in Fig.~\ref{fig:show_func_dis}d. Hence, \emph{NoDis} keeps the latent codes $\mathbf{z}$ and $\mathbf{y}$ entangled.

\subsection{Comparison to other inverse projection methods}
\label{sec:compare_previous}


We now compare our inverse projection \ourname{} to iLAMP\,\citep{Amorim2012ilamp}, RBF\,\citep{amorim2015Facinghighdimensions}, and NNinv\,\citep{espadoto2019NNinv}. We show that \ourname{} yields similar if not higher quality even without using its control mechanism (which we discuss separately in Sec.~\ref{sec:control_and_ID}).

\begin{figure}[!htb]
  \centering
  \includegraphics[width=0.9\linewidth]{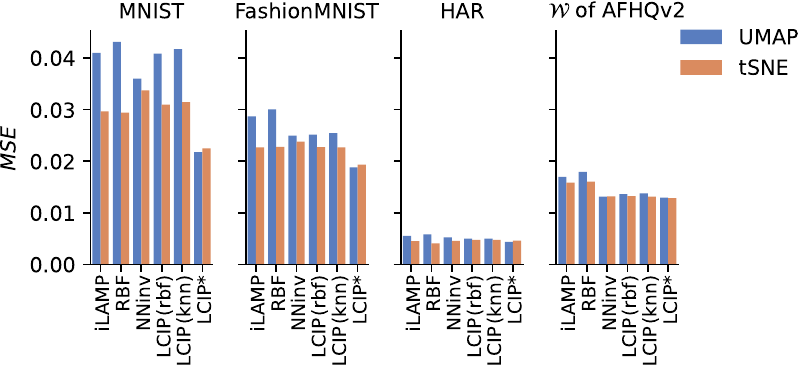}
  \caption{MSE of the studied inverse projections.}
  \label{fig:compare_mse}
  \vspace{-0.15cm}
\end{figure}

\smallskip
\noindent\textbf{Inverse projection error:} We first measure the Mean Squared Error (MSE) of the inverse projection on the test set $D_v$
\begin{equation}
  \label{eq:mse}
  MSE = \frac{1}{|D_v|} \sum_{(\mathbf{y}, \mathbf{x}) \in D_v} \|\mathbf{x} - P^{-1}(\mathbf{y}, \mathbf{z})\|^2,
\end{equation}
where $\mathbf{z}$ is ignored for iLAMP, RBF, and NNinv. For \ourname, we compute $\mathbf{z}$ at $\mathbf{y} \in \mathbb{R}^2 \setminus Y$ by smoothed RBF and weighted k-NN interpolation of the $\mathbf{z}_i$ values (see Sec.~\ref{sec:initialize_z}). We call these two interpolation variants \ourname{\,(rbf)} and \ourname{\,(knn)}, respectively. We also evaluate $\mathbf{z}_i = Enc(\mathbf{x}_i), \mathbf{x}_i \in X_v$ and denote this computation as \ourname{*}. While this is not the intended way to get $\mathbf{z}$ (as we cannot do this for \emph{any} 2D point, see Sec.~\ref{sec:initialize_z}), this gives the performance of our method if we could compute $\mathbf{z}$ exactly. 
Figure~\ref{fig:compare_mse} shows that iLAMP, RBF, NNinv, and \ourname{} have similar MSE on $D_v$. Yet, \ourname{*} has a lower MSE than the others on MNIST and Fashion-MNIST. Hence, our method can produce inverse projections with \emph{lower error} when $\mathbf{z}$ is properly provided. This quality gap between \ourname{*} and \ourname{\,(rbf)} or \ourname{\,(knn)} can be filled by interaction (see next Sec.~\ref{sec:application_afhq}).  

\smallskip
\noindent\textbf{Visual quality in gap areas:} MSE testing can only be done for points in $D_v$, \emph{i.e.}, where we have ground truth data samples for $P^{-1}$. Yet, it is precisely in the gaps between 2D projected points that one wants to use inverse projections (Sec.~\ref{sec:basics}). One way to assess quality there is to study how inversely projected samples from gap areas \emph{look} like. Ideally, we want to obtain \emph{plausible} samples which follow the overall nature of the data in a given dataset.

\begin{figure}[!ht]
  \centering
    \includegraphics[width=0.93\linewidth]{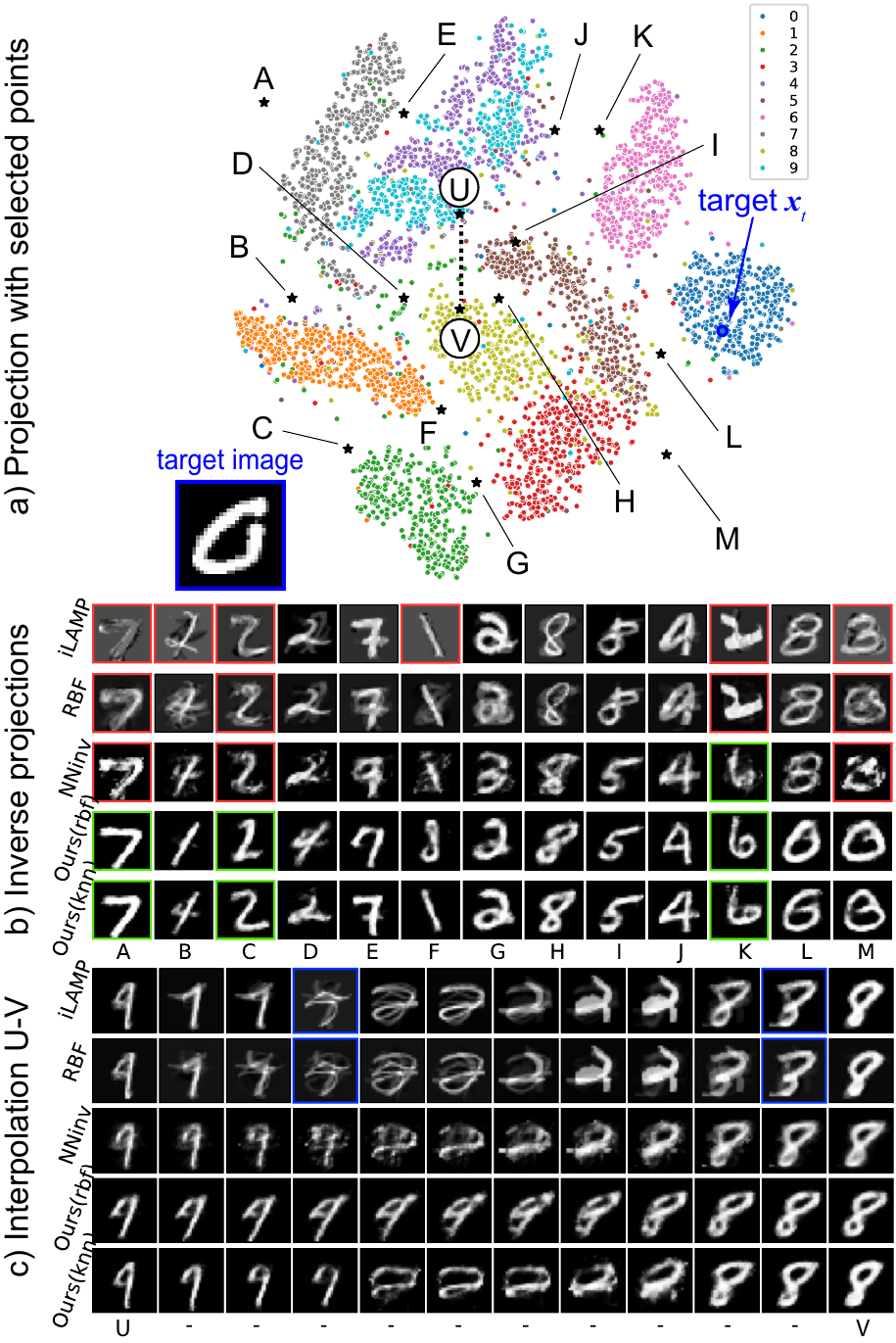}
    \caption{Visual comparison of inverse projection, MNIST dataset. (a) 13 points A-M are selected in the projection space at various distance from projected samples. (b) The inverse projections at these locations using the tested inverse projection techniques. (c) Results obtained by inversely projecting points along a line between the locations U and V in the projection.}
    \label{fig:compare_visual_mnist}  
    \vspace{-0.15cm}
\end{figure}

Figure~\ref{fig:compare_visual_mnist} shows the four studied $P^{-1}$ methods on MNIST. Images (b) show that \ourname{} produces more plausible digits than the other  methods even at locations far away from data samples (\emph{e.g.}, A, C, M). We also see that iLAMP and RBF are more sensitive to outliers. For example, consider the green point (near K, class 2) located in between the pink (class 4) and purple (class 6) clusters, while the main green cluster is down at the bottom of the projection. The inverse projection of `K' (close to the green point) is a `2' in iLAMP and RBF, while it is a `6' in NNinv and \ourname{}, so iLAMP and RBF feel the effect of this single outlier strongly whereas NNinv and \ourname{} do not.  Separately, iLAMP creates gray backgrounds (\emph{e.g.}, A,B,C,F,M) which are not only far away from the actual data distribution but also not what one would expect for MNIST. Images (c) show the inverse projections from points along a line between images U (digit 9) and V (digit 8) in the projection. iLAMP, RBF, and NNinv generate spurious shapes that do not resemble any digit during this interpolation process -- very likely this due to the few outlier points from different classes present along this line. In contrast, LCIP morphs the 9 to an 8 following, we argue, a more natural set of intermediate images. As such, if one desires an inverse projection which is less sensitive to outlier points, LCIP is preferable.

A further test of the quality of inverse projections in gap areas would check whether $P(P^{-1}(\mathbf{p})) = \mathbf{p}$ for points $\mathbf{p}$ in such areas -- that is, in our image use-case, whether images created by our inverse projection would be placed by $P$ at the 2D locations from where they were inferred. Unfortunately, we cannot perform this test in general as it requires a \emph{parametric} projection function $P$. We do not want to impose such constraints on our pipeline, which can use any projection technique, including non-parametric ones such as t-SNE, among others.

\begin{figure}[!ht]
    \centering
    \includegraphics[width=0.82\linewidth]{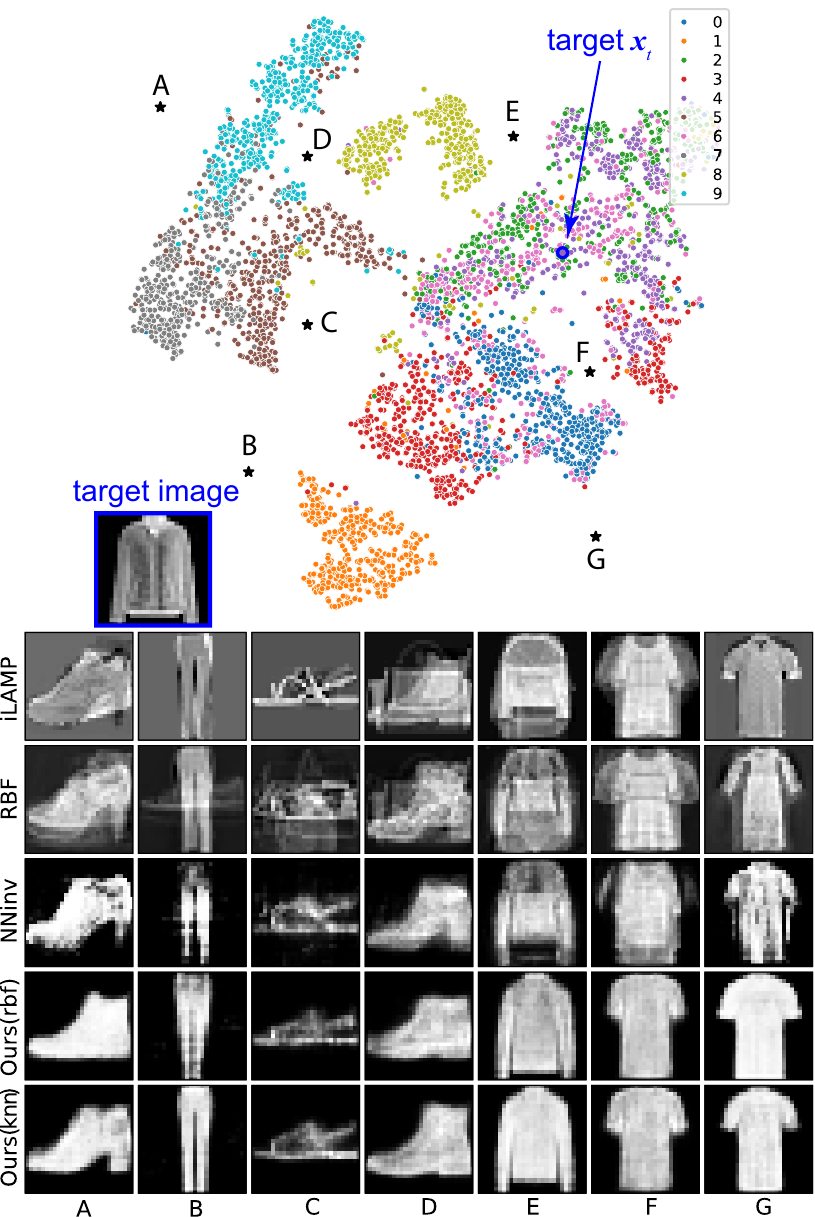} 
    \caption{Visual comparison of the inverse projection, Fashion-MNIST dataset.}
    \vspace{-0.15cm}
    \label{fig:compare_visual_fmnist}
    \vspace{-0.15cm}
\end{figure}

\begin{figure*}[!ht]
  \centering
  \includegraphics[width=\linewidth]{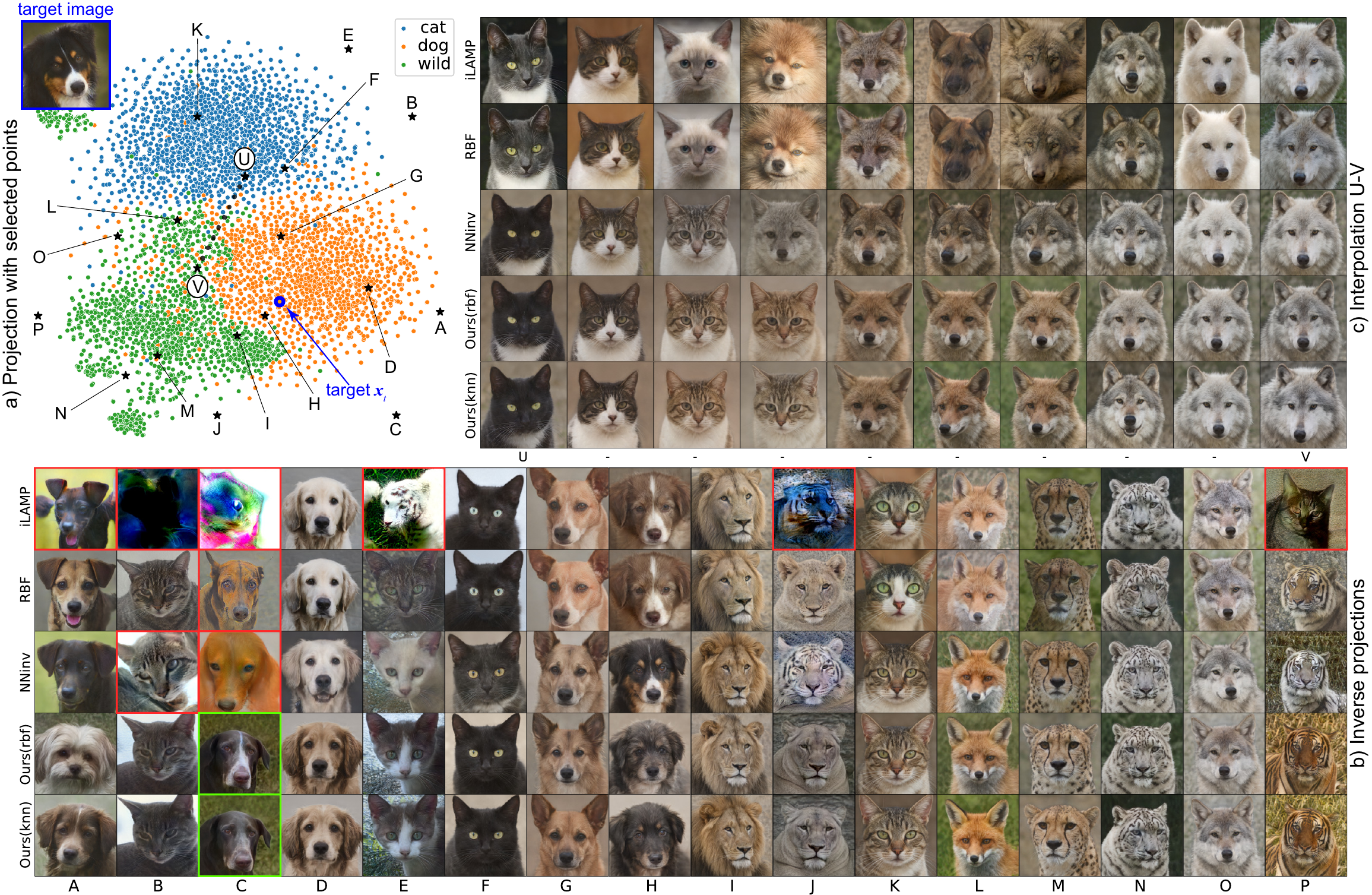}
  \caption{Visual comparison of inverse projections, AFHQv2 dataset. (a) Dataset projection with 16 selected locations both close and far away from  data samples (A-P). (b) Inverse projections at A-P created by all tested methods. (c) Inverse projections of points on a line between locations U and V in the projection.} 
  \vspace{-0.15cm}
  \label{fig:compare_visual_afhq}
\end{figure*}

On Fashion-MNIST, iLAMP and RBF mix several images in the reconstruction (Fig.~\ref{fig:compare_visual_fmnist}). For example, at point A, iLAMP and RBF mix high heels and boots; at point B, RBF mixes shoes and pants; at point F, iLAMP and RBF mix a wider and a narrower T-shirt. NNinv doesn't have this problem, but it produces jagged (\emph{e.g.}, at points A, B, C, G) or ambiguous shapes (\emph{e.g.}, at points E, F). \ourname{} keeps the reconstructed shape clear and recognizable in all cases.

On AFHQv2, all tested $P^{-1}$ methods work well when the inversely-projected 2D points are close to training samples. Once a 2D point is further away, iLAMP  becomes problematic (see A, B, C, E, J, P). In extreme cases, all methods but ours have issues. For example, at point C (Fig.~\ref{fig:compare_visual_afhq}), iLAMP produces a mass of colors with indiscernible shapes; RBF produces color distortions and a dog with only one ear; NNinv produces a dog but in a strange appearance, also with color distortions. In contrast, \ourname{} produces realistic images in all cases. 

\begin{figure}[htbp]
  \centering
  \includegraphics[width=1\linewidth]{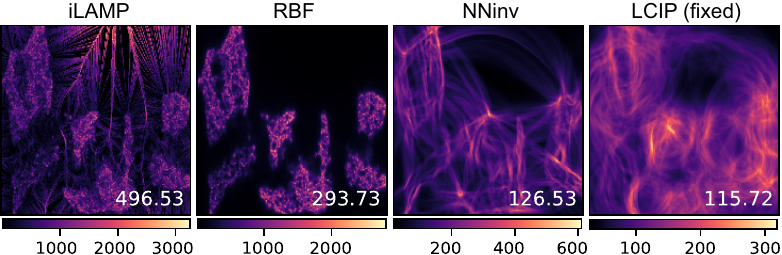}
  \caption{Gradient maps for the tested inverse projections for UMAP on MNIST.}
  \label{fig:gradient_map_mnist}
  \vspace{-0.15cm}
\end{figure}

\smallskip
\noindent\textbf{Smoothness comparison:} 
\label{sec:smoothness_compare}
We next evaluate smoothness using \emph{gradient maps}\,\citep{espadoto2021unprojection}. These are images which depict the
gradient norm $\| \nabla P^{-1} (\mathbf{p}) \|$ at every pixel $\mathbf{p}$ of the projection space. Figure~\ref{fig:gradient_map_mnist} shows these maps for the tested inverse projections on MNIST with the UMAP projection; additional datasets and projections, shown in the supplementary material, follow the same pattern. Bright colors (high gradients) show points where the inverse projection `jumps' in data space when its input moves between neighbor pixels. Such jumps are undesirable (see Sec.~\ref{sec:basics}). We see that LCIP achieves the smallest gradient norms (see color legends) for all datasets and direct projections. 

\begin{figure}[htbp!]
  \centering
  \includegraphics[width=1\linewidth]{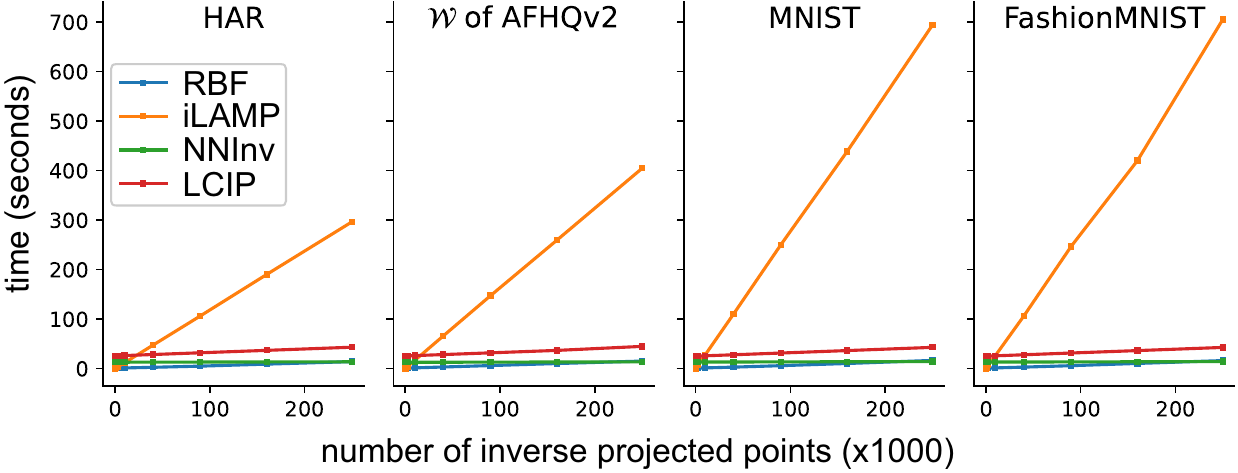}
  \caption{Inverse projection speed. The training time equals the y-intercept of the graphs. The slopes of the graphs depict how inference time depends on the number of inversely projected samples.}
  \label{fig:time_record}
    \vspace{-0.15cm}
\end{figure}

\smallskip
\noindent\textbf{Computational speed:}
\label{sec:speed}
Figure~\ref{fig:time_record} shows timings for the four studied inverse projection methods on a PC with an Intel Core i5-12400 and NVIDIA GeForce GTX 3090. The y-axis shows total time, \emph{i.e.} the (constant) training time and inference time (linear in the projected points count). All methods show similar speed except iLAMP which is significantly slower. Although \ourname{} requires slightly more training time due to its adversarial training, its slope is nearly identical to NNinv and RBF, showing the same high scalability. Separately, our evaluations show that the results of \ourname{\,(rbf)} and \ourname{\,(knn)} are very similar. Since \ourname{\,(rbf)} is theoretically smoother, we will use it in our following experiments. 

\begin{figure}[!htb]
    \centering
    \includegraphics[width=0.93\linewidth]{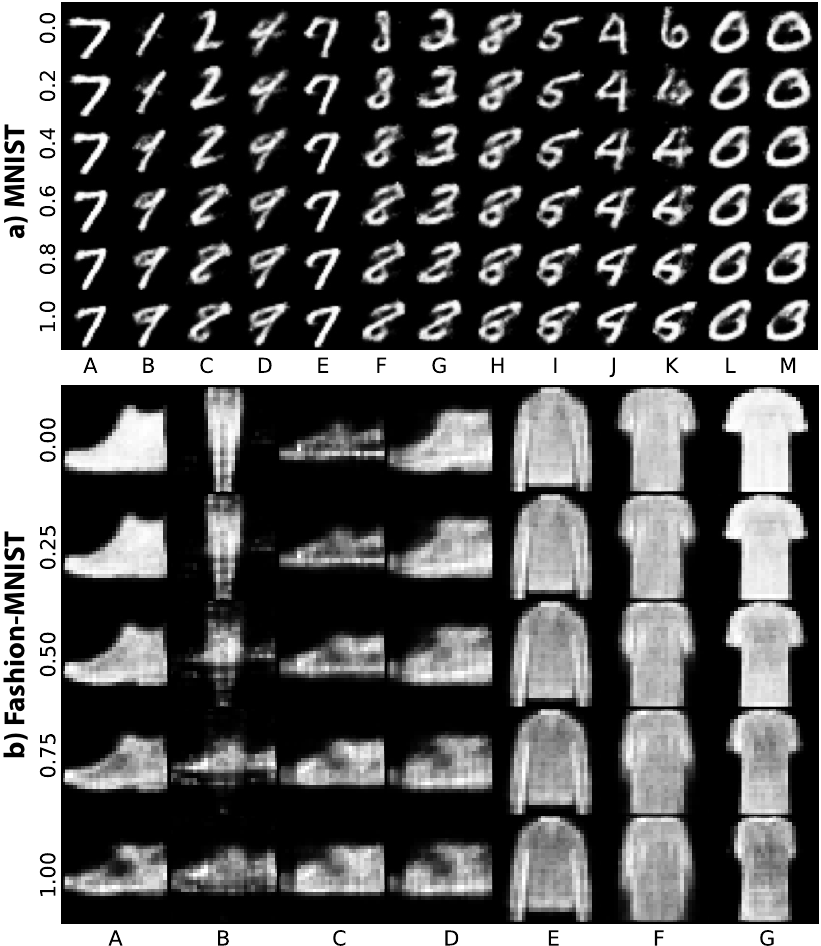}
    \caption{Controlling the inverse projection for MNIST (a) and Fashion-MNIST (b). Targets are the blue-outlined inset images in Fig.~\ref{fig:compare_visual_mnist} for (a) and Fig.~\ref{fig:compare_visual_fmnist} for (b). Rows in the two images show the effect of increasing user control $\alpha$.}
    \vspace{-0.15cm}
    \label{fig:add_z_MNIST}
\end{figure}

\begin{figure*}[!htb]
    \centering
    \vspace{-0.15cm}
    \includegraphics[width=1.0\linewidth]{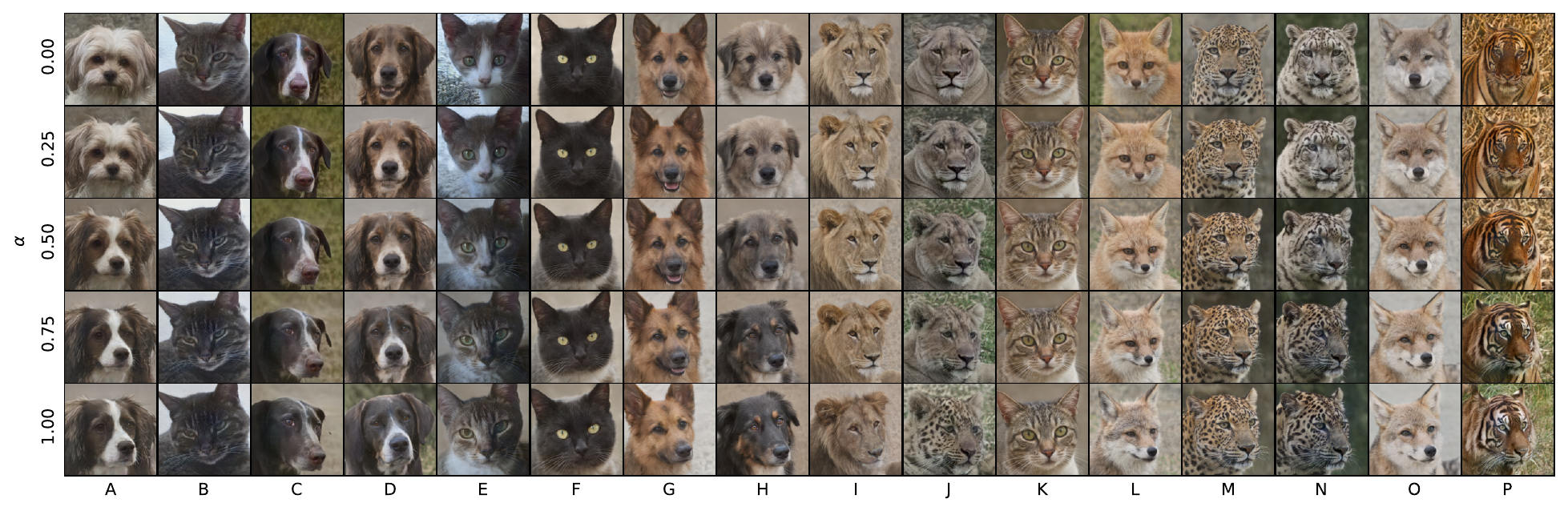}
    \caption{Controlling the inverse projection for AFHQv2. The target is the blue-outlined image in Fig.~\ref{fig:compare_visual_afhq}.}
    \label{fig:add_z_afhq}
\end{figure*}

\begin{figure}[!htb]
  \centering
  \includegraphics[width=1.0\linewidth]{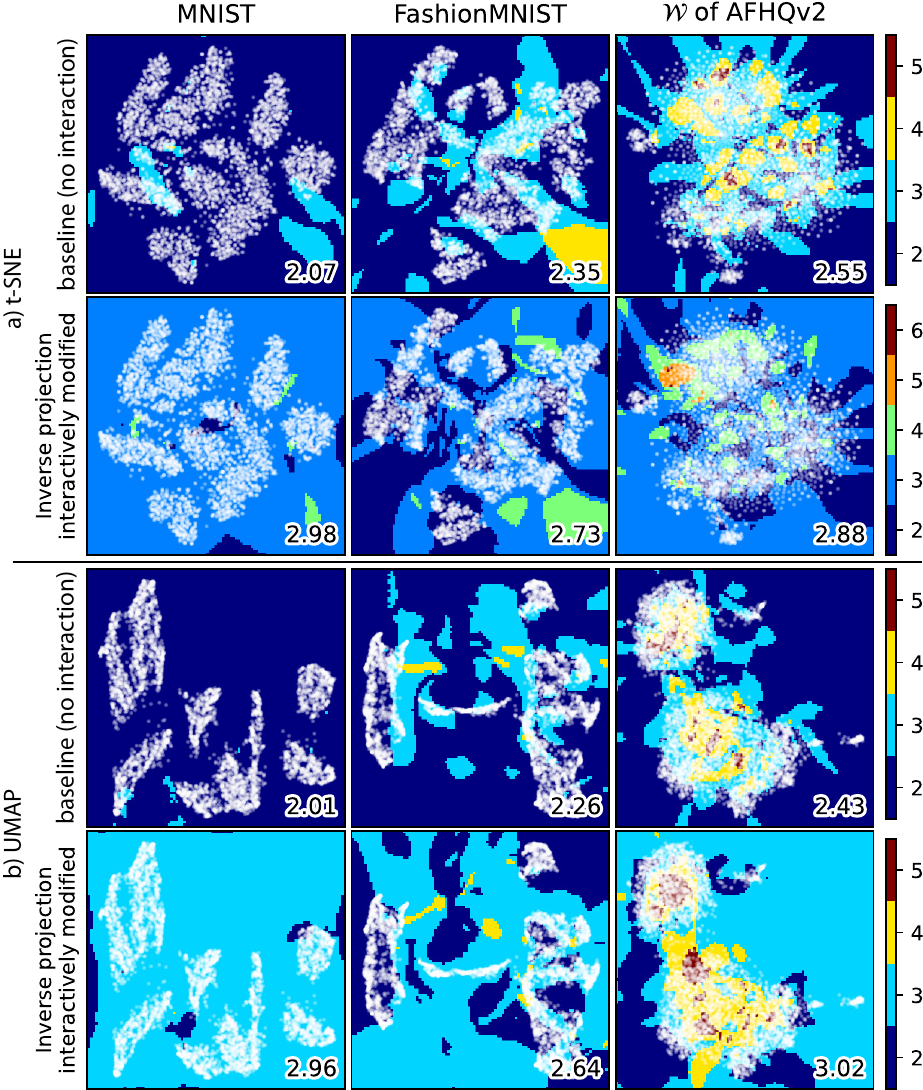}
  \caption{Intrinsic dimensionality of \ourname{} without and with interaction for different datasets and direct projections.}
  \vspace{-0.15cm}
  \label{fig:id_map_compare}
\end{figure}

\begin{figure}[!htb]
  \centering
  \includegraphics[width=1\linewidth]{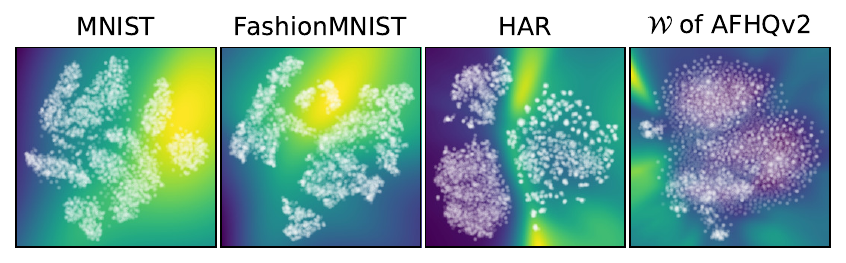}
  \caption{Normalized variance of inverse projections produced by 5000 different $\mathbf{z}$ values, t-SNE direct projection. See Eqn.~\ref{eq:variance_map} and related text.}
  \vspace{-0.15cm}
  \label{fig:var_stack_sruf}
\end{figure}

\subsection{Controllability: Going beyond a fixed surface}
\label{sec:control_and_ID}
We have shown so far that LCIP can construct a `fixed' inverse projection from a given dataset $X$ and its projection $P(X)$ with results which are comparable -- and often better -- than other existing inverse projection techniques in terms of generating plausible results in gap areas, inverse projection MSE, inverse projection smoothness, and speed.
Yet, the key feature of our method is its ability to \textbf{dynamically} control the inverse projection, which we describe next.

\smallskip
\noindent\textbf{Effects of control:} Figures~\ref{fig:add_z_MNIST}-\ref{fig:add_z_afhq} show how the inverse projection changes when their $\mathbf{z}$ values are adjusted towards selected targets in MNIST, Fashion-MNIST, and AFHQv2. Targets are marked by $\mathbf{x}_t$ (blue) in Fig.~\ref{fig:compare_visual_mnist}-\ref{fig:compare_visual_afhq} with their images shown as insets. In Figs.~\ref{fig:add_z_MNIST}-\ref{fig:add_z_afhq}, topmost rows show the selected source images $(\alpha = 0)$;  rows below show how the inverse projection `sweeps' the data space between source and target as the user increases $\alpha$. For example, when selecting an italic-like `0' digit in MNIST as target, as we increase $\alpha$, the inverse projections gradually become more italic, no matter which is the selected source image (Fig.~\ref{fig:add_z_MNIST}). We see similar changes in Fashion-MNIST, such as subtle changes in the style of shoes and shirts (Fig.~\ref{fig:add_z_MNIST} bottom). For  AFHQv2, the selected target is a dog tilting its head -- see inset image in Fig.~\ref{fig:compare_visual_afhq}. As we increase $\alpha$, all selected source animals in the inverse projection rotate their heads with similar angles (Fig.~\ref{fig:add_z_afhq}).

\smallskip
\noindent\textbf{Increased reach:} While \ourname{} generates inverse projections that go beyond a fixed surface, a key question is \emph{how much} beyond such a surface it can reach. To quantify this, we evaluate the Intrinsic Dimensionality (ID) of the inverse projection (run without interaction) using the Minimal Variance method\,\citep{tian2021Usingmultiple}. In detail, we sample a set $A$ of $100 \times 100$ pixels in the 2D projection space and map these through LCIP (without interaction) to get $Q = \{\mathbf{q}_\ell = P^{-1}(\mathbf{p}_\ell) | \mathbf{p}_\ell \in A\} \subset \mathbb{R}^n$. For each query pixel $\mathbf{p}$, we consider its inverse $\mathbf{q} = P^{-1}(\mathbf{p})$ and define the neighborhood
$S(\mathbf{p}) = \{\mathbf{q}_\ell \in Q \mid \|\mathbf{q} - \mathbf{q}_\ell\|_2 \le r\}$,
where $r = 0.1 \cdot \max_{\mathbf{q}_i, \mathbf{q}_j \in Q} \|\mathbf{q}_i - \mathbf{q}_j\|_2$ sets a fixed-radius neighborhood in data space.
We assess how close $S(\mathbf{p})$ is to an embedded surface by computing the eigenvalues $\lambda_i$, $1 \le i \le n$, of the covariance matrix of $S(\mathbf{p})$, sorted in descending order, and then evaluate 
\begin{equation}
  ID(\mathbf{p}) =
  \left|
  \left\{
  \frac{\lambda_i}{\sum_{j=1}^{n} \lambda_j} \ge \theta,\; 1 \le i \le n
  \right\}
  \right|
\label{eqn:id}
\end{equation}
with $\theta=0.05$. $ID(\mathbf{p})$ counts how many principal directions each explain at least a fraction $\theta$ of the total variance of $S$. If $ID(\mathbf{p})$ is close to $2$, it means that the inverse projection locally creates a surface around the backprojection of pixel $\mathbf{p}$ and its neighbors. We call this ID value the \emph{baseline}.

Next, we adjust LCIP by globally adding $\Delta \mathbf{z}$ values via Eqn.~\ref{eq:control_kernel}, for 50 uniformly sampled values of $\alpha \in [0, 0.2]$ (yielding $|Q| = 5\times 10^5$), and use Eqn.~\ref{eqn:id} to measure the ID of all these $5\times 10^5$ inversely-projected pixels taken together.

\begin{figure}[htbp!]
    \centering
    \includegraphics[width=0.93\linewidth]{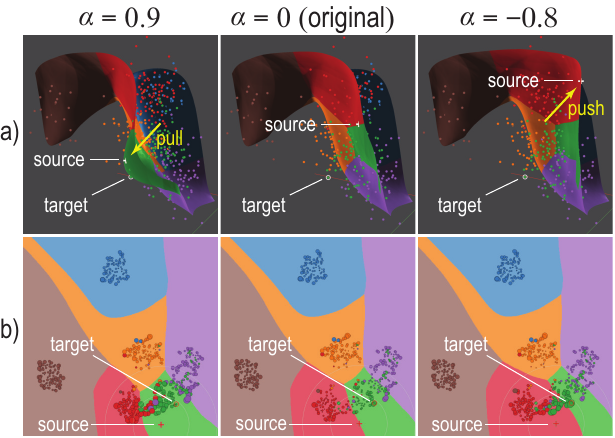}
    \caption{Inversely projected surface (top row) and corresponding decision maps (bottom row) produced by LCIP for a synthetic 3D six-class blob dataset and classified by a 3-layer fully connected deep-learning network (hidden layers of size 512/256/128). Users can vary $\alpha$ to `sweep' the 3D space with the inversely projected surface and corresponding decision maps.}
    \label{fig:dbm}
\end{figure}

Figure~\ref{fig:id_map_compare} shows that the baseline has $ID \simeq 2$ at roughly all pixels of the sampled projection space (with small higher-ID areas close to the sample points). Figures in the bottom-right corners of the images show the average $ID$ over all projection space pixels. We see that, when using LCIP without user control, LCIP creates roughly a surface embedded in $\mathbb{R}^n$, much like other inverse projection techniques\,\cite{wang2024FundamentalLimitations}. When using control, we get an ID roughly equal to 3, \emph{i.e.}, $\Delta \mathbf{z}$ `shifts' our inverse-projection surface to span a 3D space in $\mathbb{R}^n$. Some ID values of 2 are likely due to the $\mathbf{z}$ value of those areas being insensitive to the selected target, \emph{i.e.}, $\mathbf{z}_t$ and $\mathbf{z}_{p_s}$ are close in the first place. Note that we only use a \emph{single} target point here. If we used $T>1$ target samples $\mathbf{x}_t$ which span a $T$-dimensional space in $\mathbb{R}^n$, we would obtain an inverse projection of ID $\simeq T$. Figure~\ref{fig:dbm}a (discussed later in this section) shows the same effect: changing $\alpha$ pulls the inversely projected surface into areas far outside the original fixed surface, confirming the increased reach measured above. 

\smallskip
\noindent\textbf{Flexible control:} We further study the impact of using more control targets $\mathbf{x}_t$. For this, we consider a
training set $X_T$ with 5000 points. For each $\mathbf{x}_i \in X_T$, we compute its latent code $\mathbf{z}_i = Enc(\mathbf{x}_i)$, consider it as a target for our control mechanism, and consider each pixel $\mathbf{p}$ of the projection space as source point. For each such $\mathbf{p}$, we measure the variance 
\begin{equation} 
  \label{eq:variance_map}
  V(\mathbf{p}) = \frac{\sum_{j=1}^{n}Var(\{P^{-1}(\mathbf{p}, \mathbf{z}_i)\}^j)}{\sum_{j=1}^{n}Var({X_T}^j)}.
\end{equation}
over the set $\{ Dec(\mathbf{p}, \mathbf{z}_i) \}$, across all data dimensions $1 \leq j \leq n$. Higher values of $V(\mathbf{p})$ indicate that the inverse projection of $\mathbf{p}$ is more sensitive to control, and vice versa. The denominator in Eqn.~\ref{eq:variance_map} accounts for the spread of the training set $X_T$ in data space. If this spread is low, we should not expect that our inverse projection reaches far in the data space, and vice versa.

Figure~\ref{fig:var_stack_sruf} shows the results. On $\mathcal{W}$ of AFHQv2 and HAR, variance is lower near data samples and higher in gap areas, telling that inverse projections are more `nailed' when there is ground-truth data around, and more flexible when there is no ground truth nearby. On MNIST and Fashion-MNIST, the pattern seems not to be related to the distance to data samples. We believe that this is an effect of t-SNE's known tendency to compress (or stretch) point neighborhoods from data space to  projection space\,\citep{tvisne}. That is, areas showing a low normalized variance in Fig.~\ref{fig:var_stack_sruf} may actually map points which are close in data space, where our inverse projection does not have the freedom to move much; conversely, areas of high variance may map points far away in data space, where our inverse projection has more freedom to move. These maps provide users with insights on where interaction is likely to be most \emph{effective}: Interacting with source points in high-variance areas will produce inverse projections which `sweep' the data space more freely, which is the core goal of interaction; interacting with source points in low-variance areas is likely not going to discover new areas in data space.

\smallskip
\noindent\textbf{Exploring decision maps:} As Sec.~\ref{sec:related_work} mentioned, inverse projections are a key element to compute decision maps which are used to depict the behavior of trained classification models of the type $f : \mathbb{R}^n \rightarrow C$\,\cite{oliveiraSDBM2022,schulz2020DeepViewVisualizing,differentiableDBM}. A decision map colors all pixels $\mathbf{p}$ of the projection space by $f(P^{-1}(\mathbf{p}))$. As explained, existing inverse projections only depict the model's behavior on a \emph{single, fixed} surface embedded in $\mathbb{R}^n$. How the model behaves outside of this surface is left unexplored.

LCIP alleviates this limitation by allowing users to interactively change the said surface to explore specific regions in the data space close to samples of interest and in directions of interest. Figure~\ref{fig:dbm} illustrates this idea using a synthetic 3-dimensional dataset with 6 classes and a lightweight fully connected classifier (this setup mirrors the example presented in\,\cite{wang2024FundamentalLimitations}). Since $n=3$, we can directly visualize the inversely projected surface created by LCIP. Figure~\ref{fig:dbm}b shows the fixed surface and corresponding decision map for $\alpha=0$, that is, similar to other inverse projection methods. We now like to explore the behavior of the classifier \emph{outside} this surface -- more specifically, close to the sample marked \emph{source} and in the direction towards the sample marked \emph{target}. 
We first use $\alpha=0.9$ to \emph{pull} the surface from source towards target. Figure~\ref{fig:dbm}a shows how the surface locally deforms accordingly; this effectively demonstrates the effect outlined earlier in Fig.~\ref{fig:control}.
At the same time, the decision map changes to show us how the model behaves in previously unseen areas in data space that the deformed surface goes through now. Next, we set $\alpha=-0.8$ to \emph{push} the surface from source away from the target (Fig.~\ref{fig:dbm}c). This explores the model on the other side of the original fixed surface. Interactively changing $\alpha$ allows us to see the model's behavior on a \emph{dense set} of surfaces which sweep the data space around the source and in the positive/negative direction of the target. Choosing different sources and targets enriches our understanding of the visualized classifier in specific areas of interest.

%% file: sections/application.tex
\section{User control of the inverse projection}
\label{sec:application_afhq}
We now show how the control of \ourname{} works in practice using the tool shown in Fig.~\ref{fig:teaser}a. Based on how far the selected source $\mathbf{p}_s$ is from the selected target $P(\mathbf{x}_t)$ in projection space, we have in practice two types of control: target is (1) \emph{close} to, respectively (2) \emph{far away} from, the source. In (1), we expect $P^{-1}$ to gradually but \emph{fully} change towards the target; In (2), we only expect a \emph{partial} change, \emph{e.g.,} style-wise. We show next that \ourname{} yields indeed these two different effects.

\begin{figure}[!ht]
  \centering
  \vspace{-0.15cm}
  \includegraphics[width=0.95\linewidth]{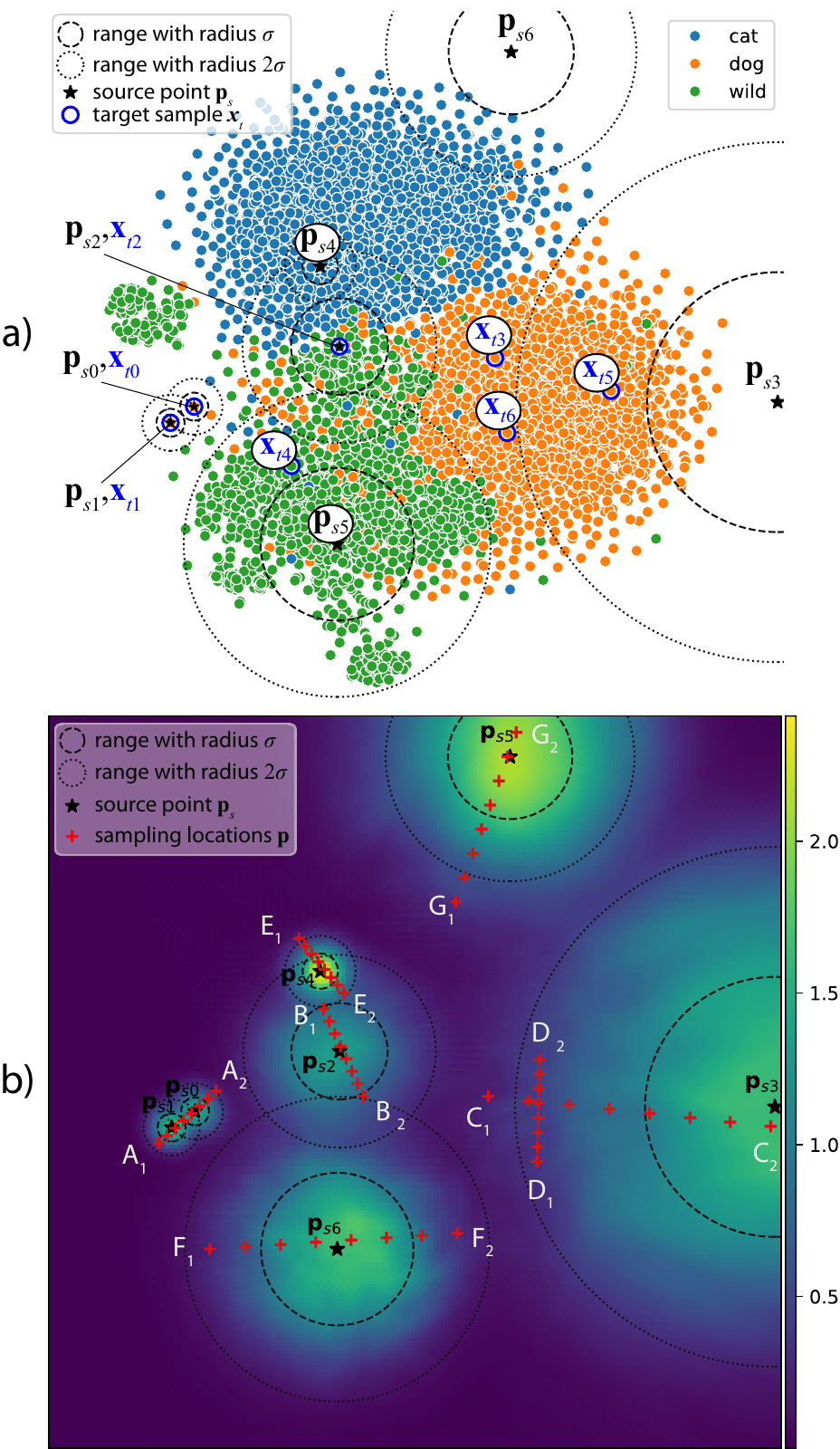}
  \caption{Close and far-away control of the inverse projection, AFHQv2 dataset. (a) 2D locations where we performed interaction.  See also Figs.~\ref{fig:control_1},\ref{fig:control_2}. (b) Assessing smoothness of controlled inverse projection at various sampling locations (\textbf{\color{red}+}). Color encodes the distance $\| \mathbf{q} -\mathbf{q}^{user} \|$ at each pixel. Figure~\ref{fig:before_after_interaction} shows the inverse projections at these sampling locations.} 
  \vspace{-0.15cm}
  \label{fig:locations_on_map}
\end{figure}

\vspace{-0.08cm}
\subsection{Local control: Target is close to source}
\label{sec:scenario_1}
We next show how our control helps with two challenges that frequently occur in inverse projection usage when the target $\mathbf{x}_t$ projects close to the source $\mathbf{p}_s$. Consider first the limit case where $P(\mathbf{x}_t) = \mathbf{p}_s$. The difference between the target $\mathbf{x}_t$ and the controlled inverse projection $Dec(\mathbf{p}_s, \mathbf{z}_{p_s})$ is fully controlled by $\mathbf{z}_{p_s}$. When $\mathbf{z}_{p_s} = Enc(\mathbf{x}_t)$, the inverse projection of $\mathbf{p}_s$ becomes $Dec(P(\mathbf{x}_t), Enc(\mathbf{x}_t)) \simeq \mathbf{x}_t$, \emph{i.e.}, the inverse projection \emph{fully} changes towards the target $\mathbf{x}_t$. 

Now consider the case where $\mathbf{x}_t$ projects close to, but not exactly at, the source $\mathbf{p}_s$. In the following, note that the choice of the target $\mathbf{x}_t$ is application dependent, \emph{i.e.}, depends on towards which data sample of special interest one wants to locally `pull' the projection.

\begin{figure*}
  \centering
  \includegraphics[width=0.6\linewidth]{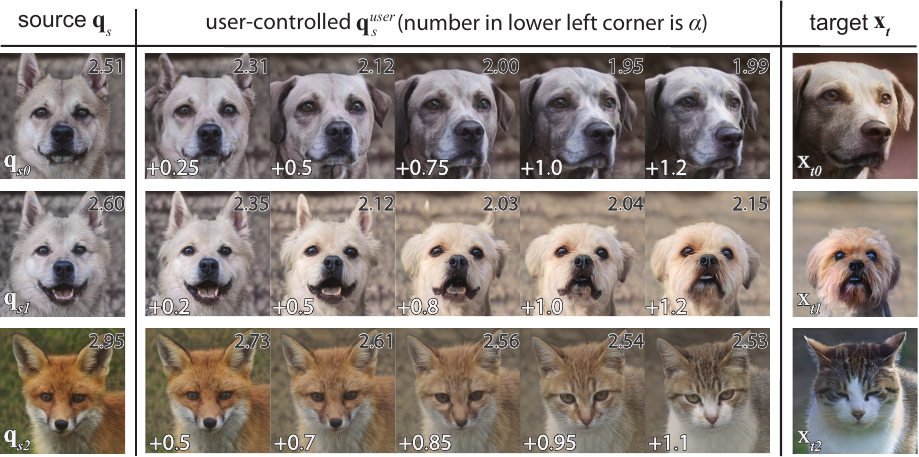}
  \caption{Local control -- adjusting the inverse projection when $P(\mathbf{x}_t) \simeq \mathbf{p}_s$. Locations of $\mathbf{q}_s$ and $P(\mathbf{x}_t)$ are shown in Fig.~\ref{fig:locations_on_map}a. Numbers in upper right corners show $\|\mathbf{x}_t - \mathbf{q}_s^{user}\|$. Numbers in lower left corners show the control parameter $\alpha$.}
  \vspace{-0.15cm}
  \label{fig:control_1}
\end{figure*}

\smallskip
\noindent\textbf{Inverse projection correction:} While good inverse projections should have a low MSE (Eqn.~\ref{eq:mse}), they do not yield $P^{-1}(P(\mathbf{x})) = \mathbf{x}$ at \emph{all} projected samples $P(\mathbf{x})$. For instance, in Fig.~\ref{fig:locations_on_map}a, $\mathbf{p}_{s0} = P(\mathbf{x}_{t0})$ and $\mathbf{p}_{s1} = P(\mathbf{x}_{t1})$ are two points at the margin of clusters. As such, their inverse projections $\mathbf{q}_{s0} = P^{-1}(\mathbf{p}_{s0})$ and $\mathbf{q}_{s1} = P^{-1}(\mathbf{p}_{s1})$ should not be strongly influenced by other points -- ideally,  $\mathbf{q}_{s0}$ and $\mathbf{q}_{s1}$ should equal the data samples $\mathbf{x}_{t0}$ and $\mathbf{x}_{t1}$ that project there. Yet, this is not the case: The inverse projection images $\mathbf{q}_{s0}$ and $\mathbf{q}_{s1}$ are not similar to the data-point images $\mathbf{x}_{t0}$, $\mathbf{x}_{t1}$ (Fig.~\ref{fig:control_1}). We can control our inverse projection to make it \emph{closer to the data}. As we add $\Delta \mathbf{z}$ to $\mathbf{z}_{p_s}$ with higher $\alpha$ values, the inverse projection $\mathbf{q}^{user}$ gradually changes towards the target $\mathbf{x}_t$ -- see Fig.~\ref{fig:control_1} top two rows. The distances $\| \mathbf{q}^{user} - \mathbf{x}_t\|$ are also reduced, reaching minima for $\alpha=1$ (Fig.~\ref{fig:control_1}).

\begin{figure*}
  \centering
  \vspace{-0.15cm}
  \includegraphics[width=0.6\linewidth]{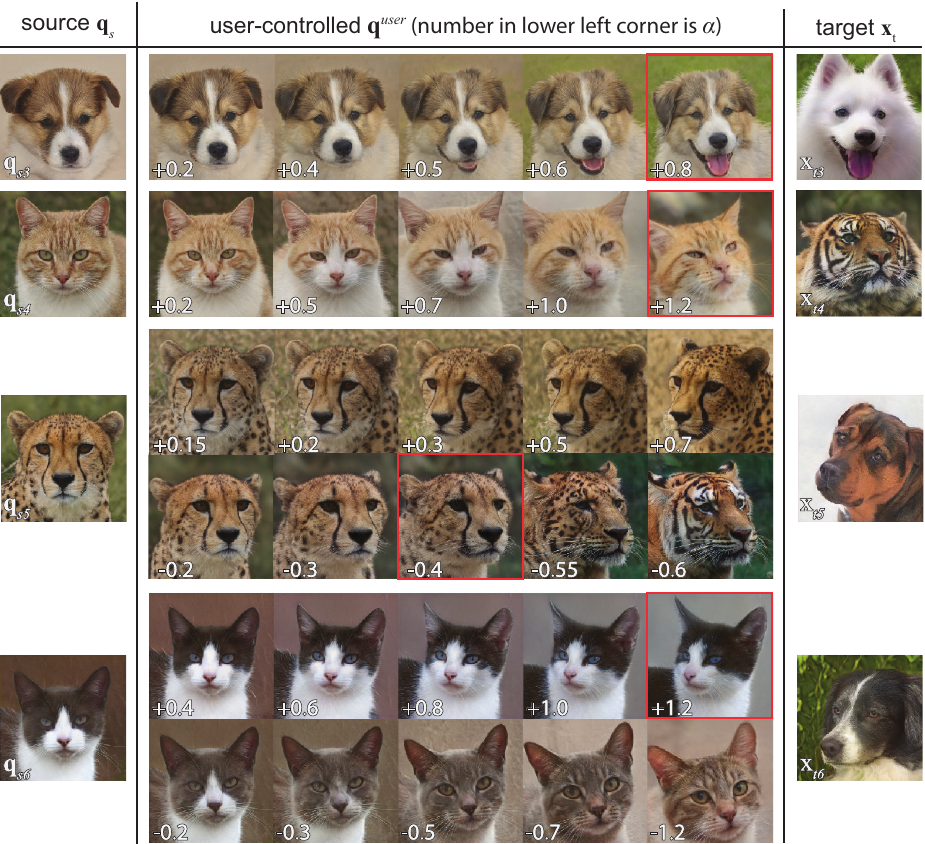}
  \caption{Far-away control -- adjusting the inverse projection when $P(\mathbf{x}_t) \not\simeq \mathbf{p}_s$. Locations of $\mathbf{q}_s$ and $P(\mathbf{x}_t)$ are shown in Fig.~\ref{fig:locations_on_map}a.}
  \label{fig:control_2}
  \vspace{-0.15cm}
\end{figure*}

\smallskip
\noindent\textbf{Overlap separation:} Overlapping points in a projection are common. More precisely, projection techniques can place samples which are (very) far apart in data space very close in the projection space -- in the limit, such points can even overlap given the finite size used to draw points in a visualization. This can be quantified by various projection quality metrics \emph{e.g.} trustworthiness\,\cite{venna2006visualizing,vennaInformationRetrieval}. For example, in Fig.~\ref{fig:locations_on_map}a, the area around point $\mathbf{p}_{s2}$ -- which corresponds to a cat (denoted $\mathbf{x}_{t2}$) -- shows overlapping samples of cats (blue) and wild animals (green). This overlap is due to the projection -- the samples are actually separable in data space. Existing inverse projection methods will map $\mathbf{p}_{s2}$ to a cat, a wild animal, or a mix of them in a \emph{fixed} and \emph{uncontrollable} way. Our method allows controlling the inverse projection of $\mathbf{p}_{s2}$ to be more like a cat or a wild animal: We see that $\mathbf{q}_{s2}$ is initially a fox (Fig.~\ref{fig:control_1}, bottom row, left). Setting the cat image $\mathbf{x}_{t2}$ as the target and $\mathbf{p}_{s2}$ as source, we see how the inverse projection gradually transforms from a fox into a cat as we increase $\alpha$ (Fig.~\ref{fig:control_1}, bottom row, columns $\mathbf{q}_s^{user}$). 

\vspace{-0.08cm}
\subsection{Far-away control: Target is far from source}
\label{sec:scenario_2}
In this scenario, the controlled $P^{-1}(\mathbf{p}_s, \mathbf{z}_{p_s})$ will not go completely toward $\mathbf{x}_t$, since $P(\mathbf{x}_t) \not\simeq \mathbf{p}_s$. Rather, only what is controlled by $\mathbf{z}$ will change. It is important to note that it is hard to tell, in general, what $\mathbf{z}$ \emph{exactly} controls as this depends on what is lost in  the projection $\mathbf{y}$, which in turns depends on the actual projection technique $P$ used \emph{and} on the dimensionality of the input data. Our specific claim is thus: Given that there \emph{is} such a loss, we (1) capture this loss and (2) allow users to `put it back' in the computation of the inverse projection -- unlike existing inverse projection techniques which behave as if this loss would not exist. 

We illustrate the above with a few examples. In our studies, we found that, for the MNIST--t-SNE combination, the digit is controlled by $\mathbf{y}$, while $\mathbf{z}$ controls the digit's style; for the AFHQv2--t-SNE combination, the type of animal faces is controlled by $\mathbf{y}$, while the animal poses are controlled by $\mathbf{z}$. Controlling $\mathbf{z}$ shapes the inverse-projected surface as desired, enabling user-controlled data generation, as illustrated next. 

Consider a source $\mathbf{p}_s$ far from any target, \emph{i.e.}, in a gap area in the projection. All existing inverse projection techniques produce surface-like structures in such areas (Sec.~\ref{sec:basics}). Our control breaks this limitation: Take point $\mathbf{p}_{s3}$ which is in a gap area (Fig.~\ref{fig:locations_on_map}a). Its inverse projection is a sad-looking puppy ($\mathbf{q}_{s3}$, Fig.~\ref{fig:control_2} top-left). We now pick as target a happy dog ($\mathbf{x}_{t3}$, Fig.~\ref{fig:locations_on_map}a, Fig.~\ref{fig:control_2} top-right). As we increase $\alpha$, the sad puppy gradually smiles and finally laughs with open mouth (Fig.~\ref{fig:control_2}, $\mathbf{q}^{user}$, top row). Note that the puppy's appearance, \emph{e.g.}, hair color and ear shapes, do not change to the target -- only its `style' changed. This control also works with source and target from different \emph{classes}. The inverse projection $\mathbf{q}_{s4}$ of $\mathbf{p}_{s4}$ is a cat looking ahead (Fig.~\ref{fig:control_2}, left column, second-top image). We choose as target a tiger looking up ($\mathbf{x}_{t4}$ in Fig.~\ref{fig:control_2}, right column, $2^{nd}$-top image). By increasing $\alpha$, the cat gradually looks up without becoming a tiger (Fig.~\ref{fig:control_2}, $\mathbf{q}^{user}$, $2^{nd}$ row).

Decreasing $\alpha$ has the opposite effect. Take a front-facing cheetah as source $\mathbf{p}_{s5}$  (Fig.~\ref{fig:control_2}, left column, $3^{rd}$ image from top); and a left-facing dog as target $\mathbf{x}_{t5}$ (Fig.~\ref{fig:control_2}, right column, $3^{rd}$ image from top). By increasing $\alpha$, the cheetah turns its head left (Fig.~\ref{fig:control_2}, $\mathbf{q}^{user}$, row 3). Decreasing $\alpha$, the cheetah turns its head right (Fig.~\ref{fig:control_2}, $\mathbf{q}^{user}$, row 4). At $\alpha \simeq -0.55$, the cheetah starts changing into a tiger, since cheetahs and tigers overlap in the projection. This $\alpha$ decrease triggers the overlap separation (Sec.~\ref{sec:scenario_1}). Finally, we choose a source $\mathbf{p}_{s6}$ in a gap area whose inverse projection is a black-and-white front-facing cat (Fig.~\ref{fig:control_2}, left column, bottom image). We set as target a left-facing dog ($\mathbf{x}_{t6}$, Fig.~\ref{fig:control_2}, right column, bottom image). Increasing $\alpha$, the cat turns its head left (Fig.~\ref{fig:control_2}, $\mathbf{q}^{user}$, row 5). Decreasing $\alpha$, the cat tilts its head right and changes fur color to brown (Fig.~\ref{fig:control_2}, $\mathbf{q}^{user}$, row 6). 

By adjusting $\alpha$, users can control subtle \emph{style} features, such as the expression or direction a subject is facing, without altering a sample's core \emph{identity}. This ability extends across different classes and also far from any projected data sample. In other words, LCIP supports tasks such as user-controlled data generation where keeping the original data's integrity, while introducing desired variations, is desired.

\subsection{Smoothness of controlled inverse projections} 
\label{sec:smoothness_check}
\begin{figure}
  \centering
  \includegraphics[width=1\linewidth]{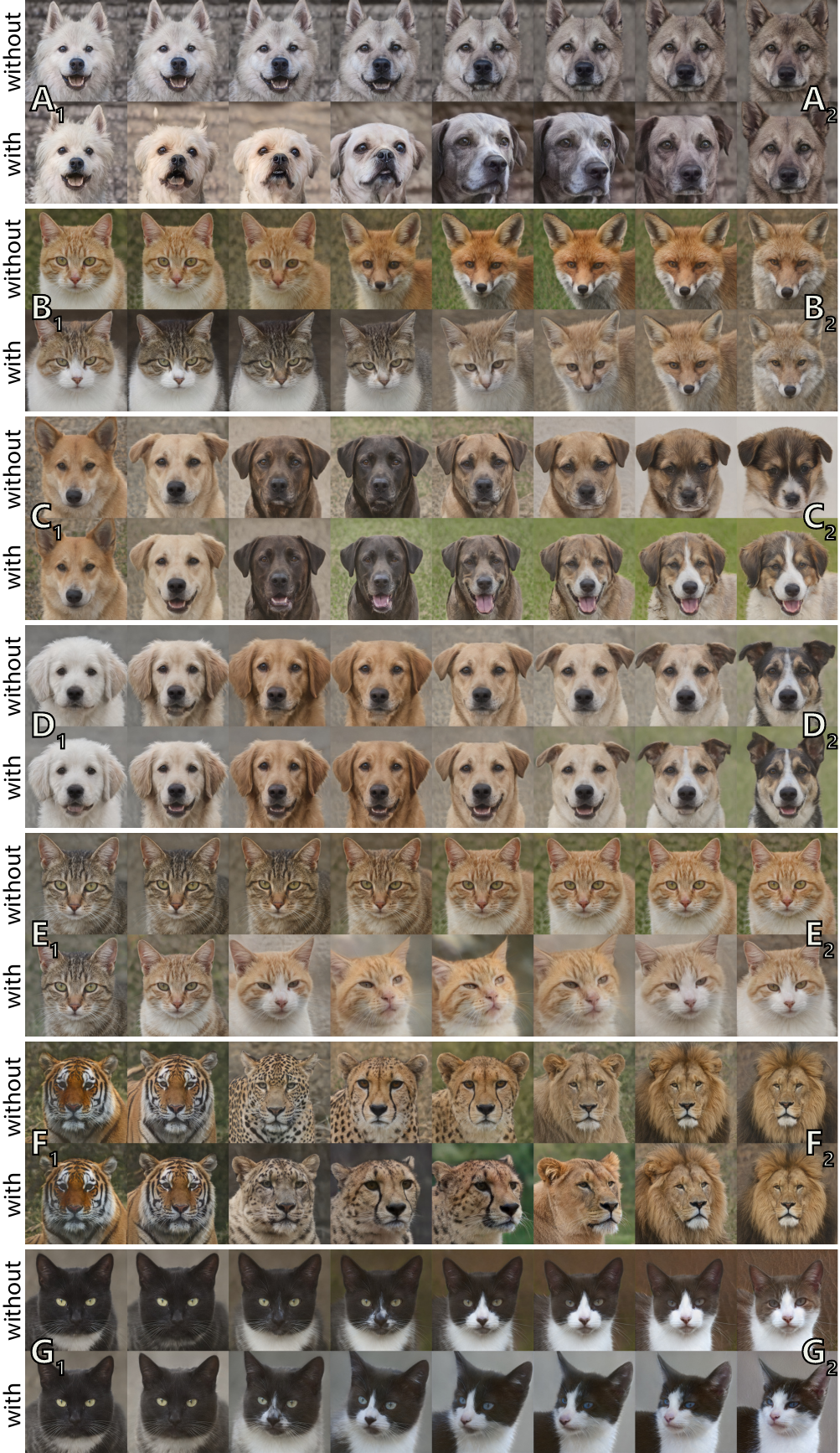}
  \caption{Smoothness with/without control: Inverse projections at 8 locations (columns) around 7 source points (rows). Images are `\textbf{\color{red}+}' marks in Fig.~\ref{fig:locations_on_map}b.}
  \label{fig:before_after_interaction}
    \vspace{-0.15cm}
\end{figure}

Section~\ref{sec:smoothness_compare} showed the smoothness of our LCIP used without user control. User control should create a smooth inverse projection since $Dec$'s input smoothly varies with $\mathbf{p}$, $\alpha$, and $\sigma$ (Eqn.~\ref{eq:control_kernel}) and since $Dec$ itself is smooth (being the neural network described in Sec.~\ref{sec:method}). 
To practically test this smoothness, we choose 7 source points $\mathbf{p}_{s0} \ldots \mathbf{p}_{s6}$ and sample several lines $A_1 A_2$ \ldots $G_1 G_2$ in projection space with 8 samples each. Figure~\ref{fig:locations_on_map}b shows the sampling locations (\textbf{\color{red}+}) and source points ($\bigstar$). We next set different $\alpha$ and $\sigma$ values for the source points and compute their inverse projections. Figure~\ref{fig:before_after_interaction} shows the inverse projections without and with control. Images in each row -- thus, over a set of linearly-spread sampling points -- smoothly change in both cases. To further confirm this, Fig.~\ref{fig:locations_on_map}b color-codes the difference between the inverse projection without and with control $\| \mathbf{q} - \mathbf{q}^{user} \|$ at each pixel. The result is a \emph{smooth} signal, with large values close to the source points and small values further away, exactly as aimed by local control (Eqn.~\ref{eq:control_kernel}). Supplementary material Fig.~2 supports this by showing gradient maps before and after user control. The smoothness of \ourname{} with user control is further confirmed by a user study (see Sec.~\ref{sec:user_study}, Task 3).
Yet, Fig.~\ref{fig:before_after_interaction} shows that some image sequences change more rapidly than others -- see \emph{e.g.} the sequence $C_1 C_2$, third row from top. This is explained by the fact that equal steps in the projection space (Fig.~\ref{fig:locations_on_map}b) may correspond to small, or larger, changes in data space due to the nonlinearity of the underlying t-SNE projection.

%% file: sections/user-study.tex
\section{User study: Evaluating LCIP in practice}
\label{sec:user_study}
We performed a user study with 15 participants (\textbf{P01}--\textbf{P15}) -- undergraduate students to postdoc researchers in infovis/visual analytics -- 3 females and 12 males, aged 24-41 years (mean: 27.3). Full details on participants and protocol are in the supplementary material. 
We used four tasks to evaluate the \emph{Uniqueness} (T1) and \emph{Fidelity} (T2) of the generated data; and \emph{Smoothness (Natural Interpolation Order)} (T3) and \emph{Controllability} (T4) of LCIP. We used AFHQv2 (Figs.~\ref{fig:teaser},~\ref{fig:locations_on_map}), a complex but still interpretable dataset.  T1-T2 follow Amorim \emph{et al.}'s evaluation of inverse projections\,\cite{amorim2015Facinghighdimensions}. For T1-T3, images were printed on paper ($5 \times 4.5$ cm) to ease comparison, assignment, and arrangement. Upon task completion, we took pictures to document the participants' arrangements. For T4, 
participants used our tool (Fig.~\ref{fig:teaser}); we gathered user feedback on controllability and usability via post-task interviews. 

\begin{figure}[t]
    \centering
    \includegraphics[width=0.95\linewidth]{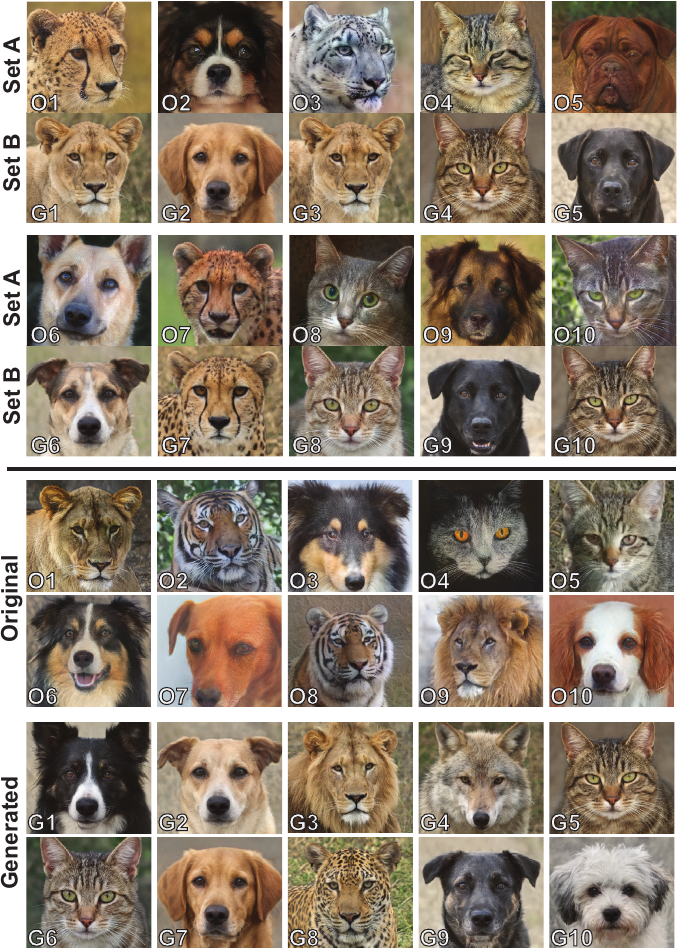}
    \caption{Top: The 10 image pairs used for Task 1: Uniqueness. Bottom: The 10 original images (O1--O10) and 10 generated images (G1--G10) used for Task 2: Fidelity. See Sec.~\ref{sec:user_study} T1,T2.}
    \label{fig:user_study_task1}    
    \vspace{-0.15cm}
\end{figure}

\smallskip
\noindent
\textbf{T1 Uniqueness:}
Participants were shown 10 images from AFHQv2 (set A) and 10 generated images (set B), see  Fig.~\ref{fig:user_study_task1} top.
For each image in B, participants were asked to choose the most similar image in A or to reject all images if none appeared similar. Several B images can be assigned to one A image. This task also enabled participants to get to know the dataset as they were always informed if images were original or generated. We selected images in A by random sampling AFHQv2. Images in B were created by \ourname{} (fixed) applied to the original image's location in the projection. The uniqueness of generated images was measured by aggregating participants' choices, where fewer associations indicated higher uniqueness.

\smallskip
\noindent
\textit{Results:} Images G6 and G7 were always assigned to their original counterparts (O6 and O7). Image G10 was assigned 11 of 15 times to its original counterpart (O10) alongside G4 and G8. For O8, most users selected O8 itself as the most similar image (for full details, see supplementary material). In their assessment, most users felt that they could easily find a similar image for most images; a few could not find one at all. 

Statistical analysis confirms this finding.
Participants completed 150 matching trials with $k=67$ correct assignments, yielding an observed accuracy $\bar{p}=0.447$.
A one-sample $z$-test against the chance level ($H_0: p = 0.1$) gives $z \approx 14.1$ ($p \ll 0.0001$), so we reject $H_0$ and support $H_1: p > 0.1$.
The 95\% confidence interval $[0.367,0.526]$ lies entirely above 0.1, so participants performed 0.347 (34.7\%) above chance.

\smallskip
\noindent
\textit{Interpretation:}
\ourname{} (fixed), \emph{i.e.}, without user control, could generate images that are assignable to a ground-truth image. However, this highly depends on the local variance of the data. This can be seen by users selecting 3--4 different images in some cases, while in others, they unanimously selected the same single image. While participants said that it was not easy to find a similar image, they assigned correct counterparts above chance; this supports the claim that LCIP can generate visually distinct images. 

\smallskip
\noindent
\textbf{T2 Fidelity:}
Participants were shown 20 images (Fig.~\ref{fig:user_study_task1} bottom) and asked to tell whether each was original or generated. 10 images were randomly selected from AFHQv2; the other 10 were created by \ourname{} (fixed), sampled based on point density in the projection to ensure a representative set. By assessing one's ability to distinguish between original and generated images, this task gauges the fidelity of \ourname{}.

\smallskip
\noindent
\textit{Results:}
Most images were classified incorrectly by most participants (overall correctness was 39.67\%). See \emph{e.g.}, image G8 (13/15 original), image O3 (13/15 generated), and image G1 (12/15 original). Results for images G3, G4, and G6 were close to a tie, with 8/15, 8/15, and 8/15 classified as original, respectively. Image O6 was correctly classified as original by most users (10/15 -- 66.7\%); image G5 was correctly classified as generated by 10/15 -- 66.7\% users. The qualitative assessment shows that differentiating between real and generated images was found hard by all participants, as subtle differences such as blurred backgrounds, small irregularities, or too-perfect symmetries were often the only clues.

Participants achieved only 39.67\% accuracy, significantly below the 50\% chance level. The 95\% confidence interval $[0.342, 0.452]$ lies entirely below 0.5. This result leads us to reject the alternative hypothesis $H_1: p > 0.5$ and indicates that participants were unable to reliably distinguish between original and generated images.

\smallskip
\noindent
\textit{Interpretation:}
Our results show that participants performed worse than random chance in classifying images as original or generated. In the qualitative assessment, users mentioned looking for artifacts in the images to tell whether they were generated or not. Some participants agreed when asked whether ``normal" imperfections in an image made them think it was generated, while the generated image generally looked clean and polished. These results show that \ourname{} can create highly plausible data which is hard to distinguish from true data.

\smallskip
\noindent
\textbf{T3 Smoothness (natural interpolation order):}
Participants were given a start and end image created using LCIP and 6 interpolated images between these. The images can be observed in the \textit{with} rows in Fig.~\ref{fig:before_after_interaction} of D, G, B, C, and F, which were selected randomly, yielding five trials for this task. The goal was to arrange the interpolated images in the correct order between the start and end images. This task was conducted across three classes: cats, dogs, and wild animals.
Participants freely arranged the paper printouts on a table to reflect their perceived natural progression. The degree to which the participants' order matched the actual interpolation sequence was used to evaluate the smoothness and understandability of the inverse projection.

\smallskip
\noindent
\textit{Results:}
Participants could recover the original interpolation order at the following rates: 11 (trial 1); 13 (trial 2); all (trial 3); all but one (trial 4); and 11 (trial 5). In the qualitative assessment, the participants said that they found it easy to find a natural interpolation order for the images. In cases with a larger change in the animal's appearance, such as fur color, the participants mentioned using other features like facial expressions or viewing directions to create a meaningful order.

For each trial $T_j$, we compared the observed number of correctly ordered intermediate images with the chance baseline $\mu_{\text{chance}}=1$ (six intermediate images give a $1/6$ success probability). One-sample $t$-tests strongly rejected $H_0: \mathbb{E}[T_j] = \mu_{\text{chance}}$ in favor of $H_1: \mathbb{E}[T_j] > \mu_{\text{chance}}$; even the weakest trial yielded $t \approx 12.9$ ($p \ll 0.001$). Hence participants achieved significantly higher scores than chance.

\smallskip
\noindent
\textit{Interpretation:}
Most users recovered the natural interpolation order created by LCIP, indicating that our approach creates visually smooth transitions between points, even in the case of \ourname{} with user control. Since users were particularly successful in trial 3, our results indicate that LCIP can create meaningful and smooth transitions between species (here, cat and fox).

\smallskip
\noindent
\textbf{T4 Controllability:}
Participants were introduced to LCIP in the context of image manipulation and engaged in an initial free exploration phase. Next, they had to perform two tasks: (1) Select an inverse sample image of a dog and modify it to make the dog smile, using source image 2852 (a white laughing Samoyed) as reference. (2) Do the same with a cat image.

\smallskip
\noindent
\textit{Results:}
All users could transfer the facial expression of image 2852 to other dogs. The interaction with the $\alpha$ slider was perceived as relatively intuitive and easy. The effect of the $\sigma$ slider was less obvious to users. When trying to transfer the facial expression from the dog to a cat image, users were in general not successful, yet noticed that other cat features, \emph{e.g.} fur or viewing direction, got more similar to the dog image.

T4 was perceived as particularly interesting. The ability to directly manipulate images and transfer characteristics such as ``sticking out the tongue'' was highlighted positively by all participants (P1 -- P15). There were difficulties in consistently transferring desired changes between different animal species, such as ``sticking out the tongue'' or emotional expressions (e.g., sadness). This was perceived as a technical limitation (P1, P4, P7, P15). Participants initially struggled to understand the meaning of parameters such as $\alpha$ and $\sigma$. However, with practice, understanding improved for over half of the participants (P2, P3, P7, P9, P11--P15). It was criticized that the parameters were insufficiently labeled and explained. Suggestions for improvement included clearer labels and more illustrative explanations of the relationship between parameters and high-dimensional spaces (P3, P7, P13, P14). The simple handling of the slider, the clear visual representation of changes, and the ability to compare projections with original images were perceived as helpful and functional (all participants except P1, P4, and P8). Some participants wished for more precise control options to specifically adjust certain image features, such as fur color, eye shape, or body posture (P9, P12--P15). The ability to manipulate projections and visually track the effects of changes were perceived as exciting and useful (P10--P14).

\smallskip
\noindent
\textit{Interpretation:}
Most T4 participants (a) had overall positive impressions of LCIP's ability to support style transfer and (b) consistently managed to obtain the desired results for the dog image (less so for the cat image). Expectedly, learning to effectively use LCIP's parameters to obtain the desired results took some effort. Importantly, we should note that this application is quite challenging and its success depends not only on LCIP's abilities but also the richness of the ground truth data -- for example, if only a few smiling cat examples exist compared to dog ones, it is inherently harder for any interpolator to transfer a smile style from a dog to a cat.

%% file: sections/disscussion.tex
\section{Discussion}
\label{sec:discussion}

We next discuss several aspects of our method.

\smallskip
\noindent\textbf{Controllability:} To our knowledge, \ourname{} is the \emph{first} user-controlled inverse projection that breaks the limitation that inverse projections land on a surface embedded in $n$D (Sec.~\ref{sec:control_and_ID}). \ourname{} allows this inversely-projected surface to be put anywhere and smoothly adjusted (Sec.~\ref{sec:smoothness_check} and \textbf{Task 3} in Sec.~\ref{sec:user_study}).

\smallskip
\noindent\textbf{Genericity:} We demonstrated LCIP with image style-transfer as this an illustrative application for inverse projections. Yet, LCIP is not limited to image data, as shown by the controllable decision maps example in Sec.~\ref{sec:control_and_ID}; nor should it be seen as \emph{competing} with dedicated image-style-transfer methods. Our method can control inverse projections created by any user-selected projection $P(X)$ of any high-dimensional dataset $X$. On the other hand, the practical application of LCIP -- or, for that matter, any other inverse projection -- to explore a high-dimensional data space works easiest when the data are directly \emph{displayable}, \emph{e.g.}, for  images or 3D shapes\,\cite{amorim2015Facinghighdimensions}. If data were abstract feature vectors obtained by \emph{e.g.} an autoencoder, or samples of a time series, LCIP (or other inverse projections) would technically work equally well but user \emph{control} of the inverse projection would be less intuitive. This is not a problem of inverse projections but of how to intuitively display high-dimensional data samples in a 2D projection space. How to do this effectively for non-visual data \emph{e.g.} audio data or general-purpose feature data is an open problem for future research. 

\smallskip
\noindent\textbf{Quality:} Our evaluations show that LCIP has higher quality in gap areas than existing inverse projection techniques (Sec.~\ref{sec:compare_previous}). Also, LCIP creates smoother-varying data samples (when its 2D input changes smoothly). The user study (Sec.~\ref{sec:user_study}) further confirmed the uniqueness, fidelity, and smoothness of LCIP. Smoothness is of key added value in user-driven applications; non-smooth inverse projections would cause difficulty and  confusion for users who aim to control how 2D points are backprojected to the data space. 

\smallskip
\noindent\textbf{Scalability:} LCIP's speed is linear in the number of inversely projected points and, in practice, similar to state-of-the-art inverse projection methods, \emph{e.g.}, NNinv (Sec.~\ref{sec:speed}).

\smallskip
\noindent\textbf{Projection:} We tested LCIP using two projection methods, t-SNE and UMAP, as these methods consistently score high projection quality metrics for datasets of varying (intrinsic) dimensionality and of different provenances\,\cite{espadoto19}, making them prevalent methods in visualization and visual analytics. This means that the 2D scatterplots that t-SNE and UMAP create allow an easier exploration by users than those from a lower-quality $P$. In turn, this makes the information \emph{lost} during projection to be lower than when using a low-quality $P$, which makes LCIP's interactive control easier. Yet, for specific datasets, other techniques than t-SNE or UMAP could yield higher quality and thus be preferable as basis for LCIP. LCIP can handle any $P$ in a fully black-box manner, so using other techniques can be directly done when desired.

\smallskip
\noindent\textbf{Ease of use:} To inversely project a source point, LCIP needs minimally only selecting that point in the 2D projection space. To modify the inverse projection, one needs also to select one target point (from the projected ones) and a `pull' factor $\alpha$  telling how much the source will change towards the target. Such operations are simple to perform in a GUI by clicking (to select points) and pulling a slider (to change $\alpha$). Users can also change how smooth the inverse projection is around a source point by changing $\sigma$ -- again, just by pulling a slider.

 \smallskip
\noindent\textbf{Limitations:} LCIP disentangles the information captured by a projection $P$ from what $P$ cannot capture, and next manipulates this information to control the inverse projection. We argued that this makes good sense technically. Yet, from a practical viewpoint, as \textbf{T4} in Sec.~\ref{sec:user_study} also shows, it is by far not clear for a user what a given $P$ would capture (and thus not under user control) and what is, thus, controllable. Our experiments showed that, for selected datasets, the projection captures the core similarity of items, while the lost information -- under user control -- mainly affects a generic attribute we called `style'. Yet, what exactly style is; how it differs from what a given projection captures; and how much style is controllable in practice, are questions that we cannot formally answer for all datasets and all projection methods. 
Separately, LCIP is limited to the convex hull of samples in $X$.  As shown in \textbf{Task 4} in Sec.~\ref{sec:user_study}, transferring the ``sticking out tongue'' style from a dog to a cat image is generally not successful, likely due to this combination falling outside of the said hull. Exploring how to make $P^{-1}$ go beyond this convex-hull space is a valid option to investigate. Finally, while our examples of style transfer on images are, we hope, convincing evidence for LCIP's abilities, more use-cases are needed to strengthen our claims of added value of controllable inverse projections.


%% file: sections/conclusion.tex
\vspace{-0.15cm}
\section{Conclusion}
We have presented an inverse projection method that allows users to break the barrier of creating two-dimensional, fixed, surfaces embedded in the data space -- a property exhibited by all inverse projection methods we are aware of. To do this, we split the information present in the high-dimensional data into (1) information captured by a projection technique and (2) a latent code, namely, information that such a technique cannot capture. Next, we allow users to interactively control this latent code and thereby generate a dynamic inverse projection which can effectively `sweep' the data space between the samples used by the direct projection it aims to invert.

Several experiments show that our method meets a number of key requirements for inverse projections: Our method is -- in absence of user control -- at least as accurate, and as computationally scalable, as state-of-the-art inverse projection techniques. When user control is added, our method enables users to specify where the inverse projection should adapt to so-called target data points, and where the underlying direct projection should purely drive it. This control is simple to perform as it involves changing two linear parameters -- an action radius and an action amount. The data points generated by our method change smoothly as the control parameters change -- which is desirable from a practical perspective. Also, we showed that our method can cover a larger area of the data space -- measured in terms of intrinsic dimensionality -- than other inverse projection methods. Last but not least, we showed that our inverse projection method creates data samples (images in our studies) that look more natural than those created by existing inverse projection methods. 

LCIP could be used next to assist user-controlled 3D shape morphing\,\cite{amorim2015Facinghighdimensions} -- its control would directly allow users to select specific target shapes to use during the morphing. Separately, LCIP can help data augmentation by \emph{e.g.} subsampling a dataset to be augmented to get the sources; hand-picking specific samples (having \emph{e.g.} different styles) as targets; and creating the desired number of augmented samples by our  interpolation. We also plan to make LCIP's control more intuitive by displaying the kind of information that is captured by our latent codes, thereby showing users what they would control when manipulating LCIP's parameters. Finally, we plan to refine LCIP's local control of decision maps  (Sec.~\ref{sec:control_and_ID}) to allow analysts to explore trained models around regions of interest in simpler, but more effective, ways.